\documentclass[aps,pre,final,10pt]{revtex4-1}
\usepackage[utf8]{inputenc}
\usepackage{amsfonts, amssymb, amsmath}
\usepackage{graphicx}
\usepackage{enumerate}
\usepackage{color}
\usepackage{booktabs}
\usepackage{url}
\usepackage{hyperref}
\usepackage{comment}
\usepackage{times}

\usepackage[normalem]{ulem}

\begin{document}

\title{The role of mobility in the dynamics of the COVID-19 epidemic in Andalusia}

\author{Z. Rapti}

\affiliation{Department of Mathematics and Carl R. Woese Institute for Genomic Biology, University of Illinois
at Urbana-Champaign}

\author{J. Cuevas-Maraver}

\affiliation{Grupo de F\'{\i}sica No Lineal, Departamento de F\'{\i}sica Aplicada I,
Universidad de Sevilla. Escuela Polit\'{e}cnica Superior, C/ Virgen de Africa, 7, 41011-Sevilla, Spain}
\affiliation{Instituto de Matem\'{a}ticas de la Universidad de Sevilla (IMUS). Edificio
Celestino Mutis. Avda. Reina Mercedes s/n, 41012-Sevilla, Spain}

\author{E. Kontou}

\affiliation{Department of Civil and Environmental Engineering, University of Illinois
at Urbana-Champaign}

\author{S. Liu}

\affiliation{Department of Civil and Environmental Engineering, University of Illinois
at Urbana-Champaign}

\author{Y. Drossinos}

\affiliation{Thermal Hydraulics \& Multiphase Flow Laboratory,  
Institue of Nuclear \& Radiological Sciences and Technology, Energy \& Safety,  \\
N.C.S.R. ``Demokritos",  GR-15341 Agia Paraskevi, Greece}
  
\author{P. G. Kevrekidis}

\affiliation{Department of Mathematics and Statistics, University of Massachusetts Amherst,
Amherst, MA 01003-4515, USA}

  \author{G.A. Kevrekidis}

\affiliation{Department of Applied Mathematics and Statistics, Johns Hopkins University, Baltimore, MD 21218, USA}

\author{M. Barmann}

\affiliation{Department of Mathematics and Statistics, University of Massachusetts Amherst,
Amherst, MA 01003-4515, USA}

\author{Q.-Y. Chen}

\affiliation{Department of Mathematics and Statistics, University of Massachusetts Amherst,
Amherst, MA 01003-4515, USA}

\begin{abstract}
Metapopulation models have been a popular tool for the study of epidemic 
spread over a network of highly populated nodes (cities, provinces, countries) 
and have been extensively used in the context of the ongoing COVID-19 pandemic.
In the present work, we revisit such a model, bearing a particular case example
in mind, namely that of the region of Andalusia in Spain during the period of 
the summer-fall of 2020 (i.e., between the first and second pandemic waves). 
Our aim is to consider the possibility of  incorporation of
mobility across the province nodes focusing on mobile-phone time
dependent data,
but also discussing the comparison for our case example with a gravity  model, 
as well as  with the dynamics in the absence of mobility. 
Our main finding is that mobility is key towards a quantitative understanding 
of the emergence of the second
wave of the pandemic and that the most accurate way to capture it involves
dynamic (rather than static) inclusion of time-dependent mobility matrices
based on cell-phone data. Alternatives
bearing no mobility 
are unable to capture the trends 
revealed by the data in the context of the metapopulation model considered
herein. 
\end{abstract}

\date{\today}

\maketitle

\section{Introduction}
\label{sec:intro}

Human mobility has played an indisputable role in COVID-19 dynamics
\cite{chinazzi2020, kraemer2020} with as many as  $86\%$ of global cases
having been imported from Wuhan, the original location of the pandemic.
Studies of the epidemic in China have shown that in the early stages thereof,
the probability of an outbreak was correlated with the frequency of imported cases from
Wuhan \cite{kraemer2020}. The trajectory of the epidemic over a similar time period
was also studied in \cite{li2020} using an SEIR (Susceptible-Exposed-Infected-Recovered)
stochastic metapopulation model, where it
was determined that undocumented infections played a crucial role in the rapid spread of
the epidemic. These early and high profile studies render it clear that a 
systematic consideration of such mobility aspects of the pandemic and of theoretical
models thereof is a key ingredient towards appreciating its potential
for spreading across countries and regions. 

Continuing along this vein,
in a subsequent study \cite{galvani2020}, the probability of case importations
to countries having airports with direct flights to and from mainland China was estimated.
It was assumed that the probability of importation is proportional
to the number  of airports in the country with direct connections to mainland  China. 
With the implementation of Wuhan's travel ban and the subsequent international travel restrictions, Ref.~\cite{chinazzi2020}
analyzed the effect of quarantine measures on local, national, and international pandemic spread. Even
though the spread of the virus could only be delayed in the Chinese mainland, the mitigation of the transmission would be notable around the world.
Modeling the course of the epidemic in other countries such as England and
Wales \cite{danon2021}, also incorporated daily commuting  as an important factor in the
spread of the disease. It was assumed that infectious hosts may infect others
both at home during  the night and away during the day in the span of
a day's cycle.
A study evaluating confinement and other mitigation measures in Spain \cite{arenas2020}
used workforce mobility as a proxy for confinement. For Brazil, commuter and airline
data were used to calibrate a stochastic epidemic model \cite{costa2020}.
The model was used to investigate
the spatial spread of  the disease at various geographical scales (ranging from
municipalities to states). It should be clear that these are only some
select studies within a continuously expanding large volume of literature, which
has now also been reviewed, e.g., in~\cite{review_meta} (see also
earlier reviews such as~\cite{chen2014modeling, mccallum2001should}). 

Of course, such models have a time-honored history in earlier instances
of disease-spread modeling. For example, the Global Epidemic and Mobility (GLEAM) team
integrates real-world pandemic transmission models with mobility data, including airline transportation network flows, ground
mobility flows, and sociodemographic features, to capture spatiotemporal connections between
mobility and an epidemic's spread \cite{chinazzi2020, balcan2009}.
A model for influenza in the US \cite{pei2018}, accounted for  both
daily commuting and random travels between states. One of the main findings there was that
the metapopulation model more accurately predicts the onset, peak timing and
intensity than models only accounting for specific locations.  A study of long-term
influenza patterns in the US \cite{viboud2006} used mortality data and
the gravity model,
whereby population flows between nodes
of the metapolulation network are determined by considerations akin to Newton's law
of gravity, to study the spreading of influenza across states.  References~\cite{balcan2009, zipf1946}, also showed correlation between infection spread and
human movements.
Theoretical metapopulation studies, where the travel rates are given
by the gravity model, also exist \cite{brockmann2011}. Other works use the
rates at which hosts leave and return to their permanent locations to infer the
coupling strengths in their ODE model \cite{keeling2002}. In earlier
work, the bubonic plague epidemic was modeled using a similar approach
\cite{keeling2000}, with adjacent metapopulations on a lattice being coupled with
rates chosen to fit historical data.

Studies looking at human mobility under lenses that go beyond gravity models also exist.
One such is the radiation model \cite{simini2012}, which is based on the  assumption
that population density dictates employment opportunities, so when density is low,
commuters need to travel longer distances. Hence, the predicted flux depends on the origin and
destination populations and on the population of the region surrounding the origin location.
More  recently, a new mobility law \cite{schlapfer2021} has been proposed showing that
the number of visitors to any location is proportional to the inverse square of the product of
the frequency of visits and distance traveled. This law has been applied in the context of urban
mobility (within-city mobility), where it has shown a remarkable agreement with data.

When traffic data are available, often these are leveraged using entropy maximization techniques
\cite{gomez2019, van1980most}, aiming to reconstruct origin-destination matrices
\cite{willumsen1981} describing human mobility among various locations. However, 
in more recent considerations where
mobile phone data are available, these have been found to more accurately represent the actual movements of people
\cite{tizzoni2014, wesolowski2016}. During the first and second
COVID-19 pandemic waves, in the US \cite{badr2020association,
  glaeser2022},  Japan \cite{yabe2020non}, and in
China~\cite{chinazzi2020}, among others, mobile
phone location data were utilized to explore the effects of mobility to the reported cases reduction.

We should also note in passing that other approaches to examining
the spatial spread of COVID-19 have also been deployed, including, e.g., 
models based on partial differential equations~\cite{kevrekidis2021, mammeri2020, viguerie2021}. These modeling efforts
take into account local population density by modifying the transmission coefficients
accordingly \cite{kevrekidis2021} (compared to an ordinary differential equations model),
emphasize the importance of inflows from neighboring regions \cite{viguerie2021},  and
utilize time-varying diffusion coefficients to account for the effect of mitigation
measures \cite{kevrekidis2021, mammeri2020}.

In the present work, we wish to explore some of the practical challenges
of applying a metapopulation model to a concrete region during the COVID-19
pandemic, and also when attempting to systematically compare model results
with existing data. 
In line with our earlier studies~\cite{cuevas2021}, we bring to bear 
an epidemic model that accounts for both symptomatic and asymptomatic
infections and includes appropriate recovered compartments, as well as a compartment
for the fatalities, since the latter appears to be the most accurate dataset~\cite{holmdahl2020}.
However, since we have examined already aspects of the identifiability of
such models, as well as their usefulness in the context of age-structured
populations~\cite{cuevas2021}, we do not focus on such aspects herein. Instead, our emphasis
is on the availability of different approaches,
to couple the nodes of such a model
into a network pattern for a metapopulation description of a region of
interest. In that vein, we compare and contrast the findings of an 
implementation neglecting the mobility between provinces, with one
incorporating it. When incorporating such mobility traits, we comment
on our attempts to do so, based on ``standard'' techniques such as those
stemming from gravity models or transportation-based origin-destination matrices.

Our case example of interest is the region of Andalusia in Spain for numerous reasons,
including the familiarity of our group with the region (aiding an understanding
of the observed mobility patterns and, e.g., their seasonal variation). 
A significant feature facilitating and enabling our study is
a large-scale data
analysis of  the Transportation Ministry of the Spanish Government \cite{MITMA}
that provides time-resolved mobility data across the 
provinces within this region and hence a dynamic incorporation of the
relevant patterns based on an ``as accurate as possible'' characterization
of the mobility within the area of interest.
We calibrate the model using fatality data from Andalusia \cite{CNE}, focusing 
on the summer and early fall period  of 2020 (i.e., from around the end of the
first and the beginning of the second pandemic wave).
During this period, mitigation measures were
relatively relaxed and mobility among provinces
was high due to summer vacations and later due to higher education-related
relocation. We find that  we are unable to obtain a quantitative 
match with the observed data in each province (and hence Andalusia as whole)
without mobility 
---or with static patterns of mobility produced by some
of the above mentioned ``standard'' techniques---.
Instead, our most accurate
quantitative description of the observations stems from the incorporation
of the above described ``dynamic mobility'', as obtained from the time dependent
mobile-phone data in \cite{MITMA}.

Our presentation is structured as follows. In Section 2, we present the
model, including the relevant metapopulation network considerations.
We also show how mobility matrices, an input for the metapopulation model,
that are obtained from different data sets compare.
In Section 3, we present our results, including the parameter fitting approach
used and the comparison with the existing data for COVID-19 fatalities in each
of the Andalusian provinces. 
Finally, in Section 4, we present our conclusions
and a discussion towards future steps within these classes of models.

\section{Modeling Framework}
\subsection{Epidemic model for each node}
\label{sec:node}

\begin{figure}[!ht]
\begin{center}
\includegraphics[width=.9\textwidth]{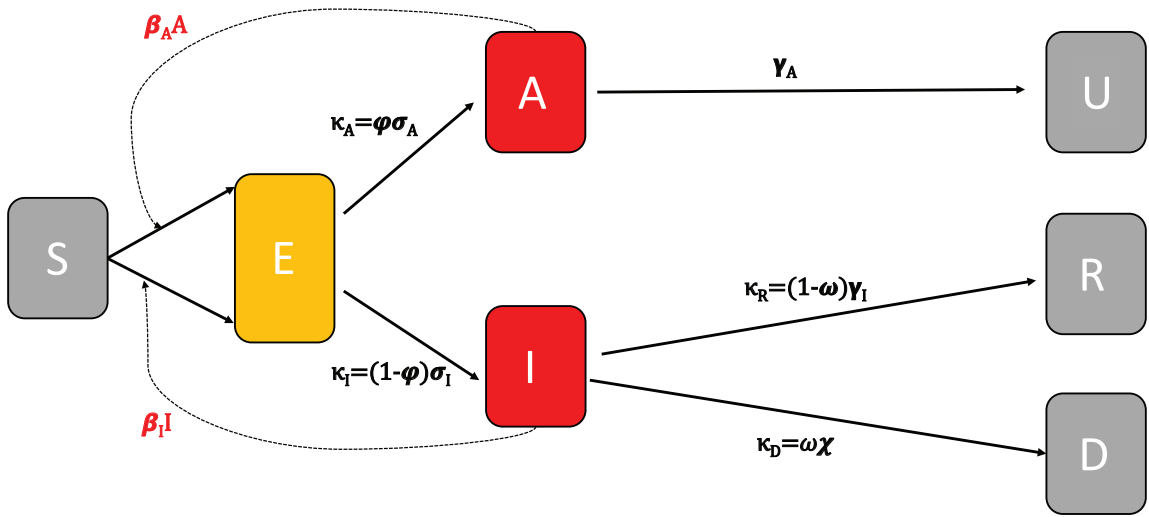}
\end{center}
\caption{Schematic diagram of the Susceptible-Exposed-Asymptomatic-Infected-Recovered (SEAIR) model for each metapopulation node.}
\label{fig:model}
\end{figure}

In the ordinary differential equation  (ODE) model that we put forth (a slight variant of the ones previously considered,
e.g., in~\cite{cuevas2021, kevrekidis2021}), 
for each node there are seven compartments. Susceptible individuals, $S$,
become exposed (latently infected, not infectious yet), $E$, after contact with either
asymptomatically infectious hosts, $A$, or symptomatically infectious hosts, $I$.
Recall that the importance of asymptomatically induced transmission, especially
in the context of COVID-19 has been argued in numerous studies~\cite{bhatta2020, review_meta}.
We assume standard incidence $\beta_{A, I}/(N-D)$,
where $N-D = S +E +A +I +U +R$ is the total living population.
However, as $D$ is quite small in most cases, from now on it will be ignored compared to $N$,
namely we will set the incidence to $\beta_{A, I}/N$, where the
transmission coefficient $\beta_{A, I}$ can be assumed constant over the considered periods of time. We have  selected the
time interval under consideration as one involving high mobility without 
changes in mitigation measures, so as to reflect more clearly the genuine role of transportation effects in the model results.

Once in the exposed class $E$, a fraction of hosts $\phi$ never develop symptoms and moves
into the asymptomatically infectious class $A$ at a rate $\sigma_A$.
Asymptomatic hosts are assumed to recover at an average rate $\gamma_A$ and move into
the recovered compartment $U$. The remaining exposed host fraction
$1-\phi$ develops symptoms
at a rate $\sigma_I$ and these individuals 
move into the symptomatically infectious class $I$.
A fraction $\omega$ of symptomatic hosts die at an average rate $\chi$ and the remaining
fraction $1-\omega$ recovers at a rate $\gamma_I$.
A schematic diagram of the above description is shown in
Figure~\ref{fig:model}. The relevant equations governing the spreading of the epidemic 
read:
\begin{align}
S' &=  - \beta_{A} S \frac{A}{N} -  \beta_{I} S \frac{I}{N}
\label{eq:sus}\\
E' &= \beta_{A} S  \frac{A}{N}+  \beta_{I} S \frac{I}{N}  - (\kappa_A + \kappa_I) E
\label{eq:exp}\\
A' &= \kappa_A E - \gamma_A A
\label{eq:asy}\\
I' &= \kappa_I E - (\kappa_R + \kappa_D)I
\label{eq:sym}\\
U' &= \gamma_A A
\label{eq:undet}\\
R' &=   \kappa_R I
\label{eq:rec}\\
D' &=  \kappa_D I,
\label{eq:fat}
\end{align}
where we set
\begin{align}
    \kappa_A = \phi \sigma_A, ~~~  \kappa_I = (1-\phi) \sigma_I,
    ~~~\kappa_R= (1-\omega) \gamma_I, ~~~\kappa_D = \omega \chi.
\end{align}
In what follows, in order to reduce parameter redundancy in the model,
we fit the
following seven parameters and parameter combinations
\begin{align}
\beta_A,~\beta_I,~\kappa_A = \phi \sigma_A, ~\kappa_I = (1-\phi)\sigma_I, ~
\kappa_R = (1-\omega) \gamma_I, ~\kappa_D= \omega \chi, ~\gamma_A
\label{eq:params}
\end{align}
This version of the model will be used when considering the fatalities
within Andalusia's provinces but without any (mobility-induced)
coupling between them and when considering the entire Andalusia 
(no metapopulation).

\subsection{Metapopulation model}
\label{sec:metapop}

We are implementing a coupling between the different
provinces in line with~\cite{brockmann2011}.
Namely, we assume that individuals are indistinguishable and travel 
from node $i$ to
node $j$ with some rate given by human mobility data, without assigning any base
location to them. Hence, individuals in node $i$ are instantaneously
assigned to node $j$ upon arrival, regardless of their prior node (no memory).
The same
individual, may change multiple nodes, in principle, 
within the model.
Connections between the nodes
depend on the mobility flow of susceptible $S$,
exposed $E$, and infectious hosts,  $A$ and $I$.
To avoid a highly complicated model, we do not incorporate
terms  such as $S_i A_j$, $S_i I_j$ in the equations., i.e.,
we assume that the primary source of infection is through interactions
of susceptible with infectious individuals within each node (no
direct long-range transmission).

The metapopulation model assumes the following form:
\begin{align}
& S_i' =  - \beta_{AS} S_i \frac{A_i}{N_i} -  \beta_{IS} S_i \frac{I_i}{N_i} +\theta \left( \sum_j M_{ij} \frac{S_j}{N_j} - \sum_j M_{ji} \frac{S_i}{N_i} \right)
\label{eq:susm}\\
& E_i' = \beta_{AS} S_i \frac{A_i}{N_i} +
\beta_{IS} S_i \frac{I_i}{N_i} - (\kappa_A +\kappa_I) E_i + \nonumber  \\
&\theta
\left( \sum_j M_{ij} \frac{E_j}{N_j} - \sum_j M_{ji} \frac{E_i}{N_i} \right)
\label{eq:expm}\\
& A_i' = \kappa_A E_i - \gamma_A A_i + \theta
\left( \sum_j M_{ij} \frac{A_j}{N_j} -  \sum_j M_{ji} \frac{A_i}{N_i} \right)
\label{eq:asym}\\
& I_i' = \kappa_I E_i - (\kappa_R + \kappa_D) I_i + \theta
\left( \sum_j M_{ij} \frac{I_j}{N_j} -  \sum_j M_{ji} \frac{I_i}{N_i} \right)
\label{eq:symm}\\
& U_i' = \gamma_A A_i
\label{eq:undetm}\\
& R_i' = \kappa_R I_i
\label{eq:recm}\\
& D_i' = \kappa_D I_i
\label{eq:fatm}
\\
& N_i' =  \theta \left( \sum_j M_{ij} - \sum_j M_{ji} \right)
\label{eq:totalm}
\end{align}
The last equation shows how the population of node $i$ is updated over time.
Our model is along the lines of \cite{li2020, pei2018}.
If
mobility is ignored by setting $\theta=0$, the
total population within each node $N_i$ is conserved.
Otherwise, solely the total population over all provinces is conserved.
$M_{ij}$ is the daily rate of people traveling from $j$ to $i$.
Then, one multiplies this rate with the proportion of $S, E, A, I$
in the total node population $N_j$.
This can be interpreted as the probability of
an individual from these four classes traveling if we  choose randomly from $N_j$.
Symptomatically infectious individuals $I$ are assumed to be able to move, but not $U$ or $R$.
In any event, the latter two do not affect further dynamics in the network as they are terminal nodes of the model. 

We note that based on the mobility data \cite{MITMA}, the network of the eight Andalusian
provinces is a complete graph and the population flows $M_{ij}$ are time-dependent. 
In the following subsection we discuss how we determined
the daily movement rates, i.e., the population flows,  between two network nodes,
and alternative ways to determine them if mobile-phone records are not available.

\subsection{Human mobility estimation}

\begin{figure}[!ht]
\begin{center}
\resizebox{0.75\textwidth}{!}{
\begin{tabular}{cc}
\includegraphics[width=.4\textwidth]{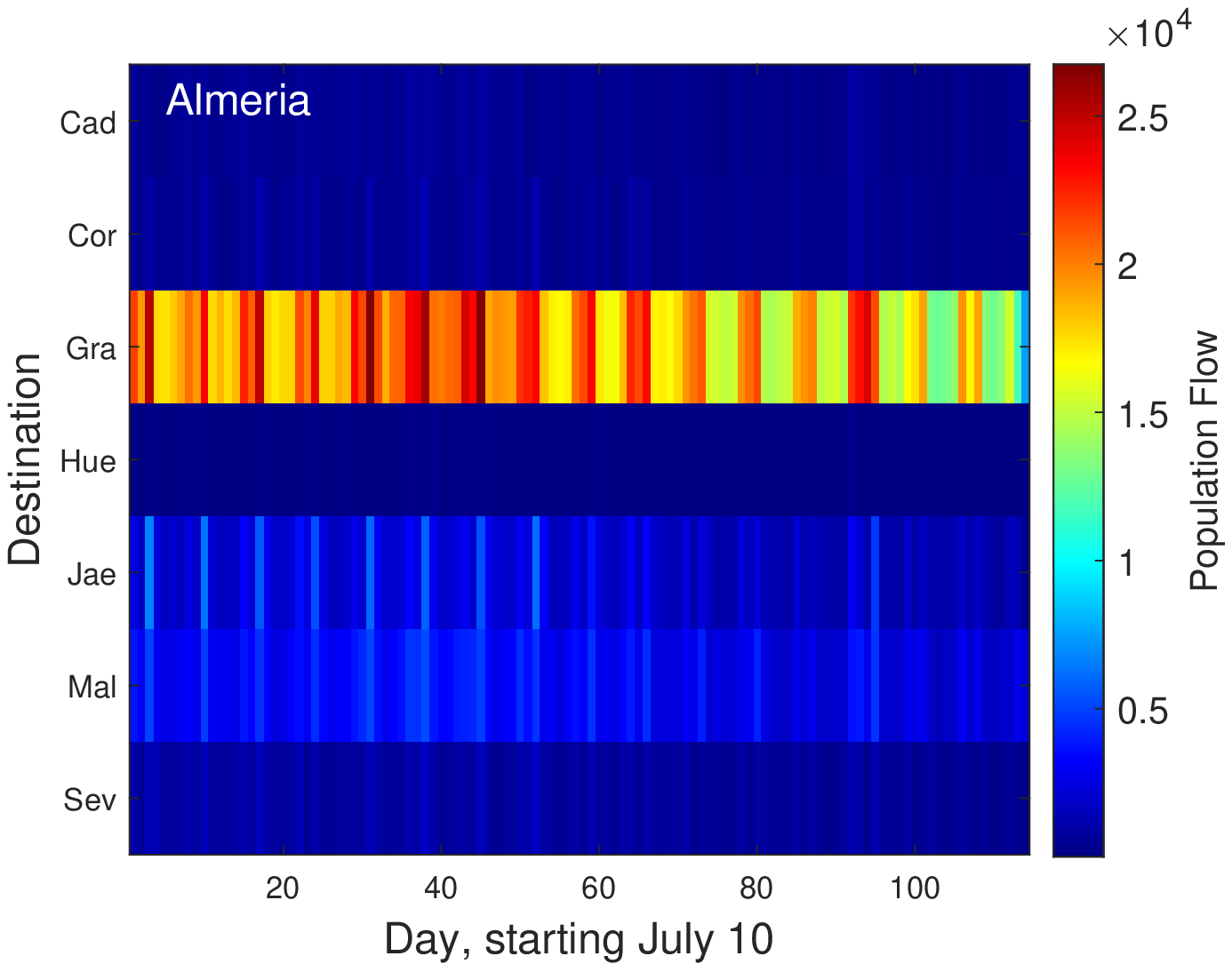} &
\includegraphics[width=.4\textwidth]{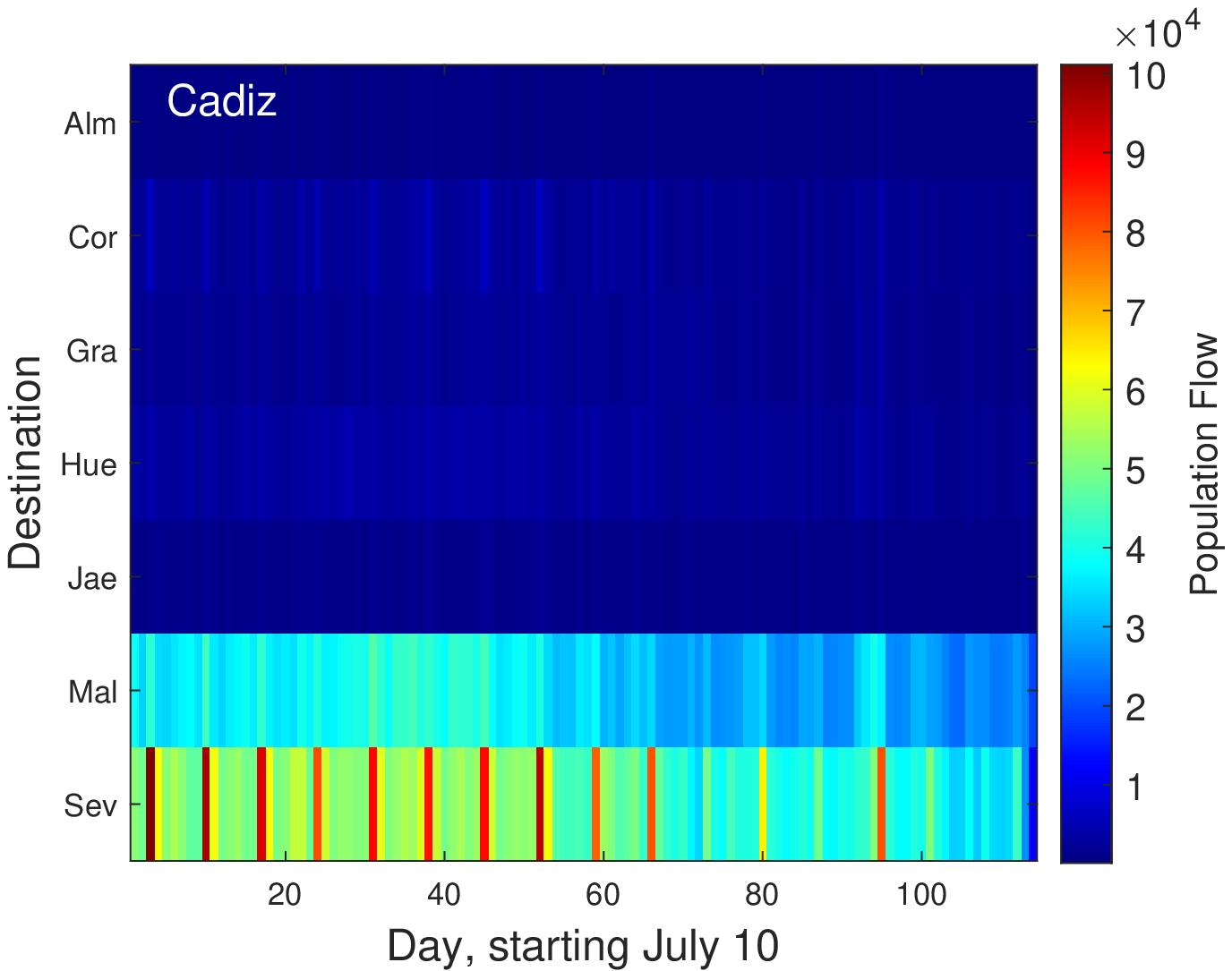} \\
\includegraphics[width=.4\textwidth]{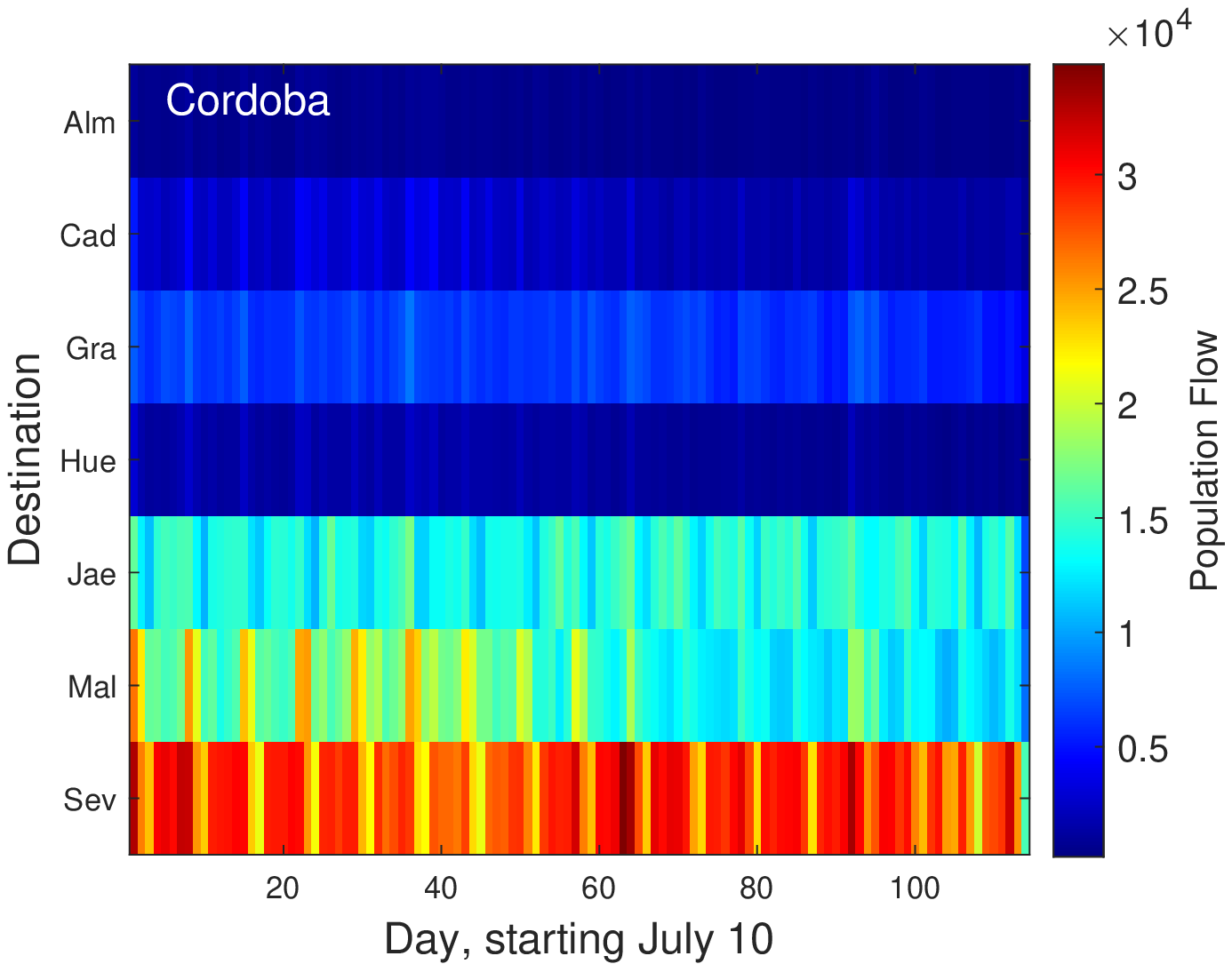} &
\includegraphics[width=.4\textwidth]{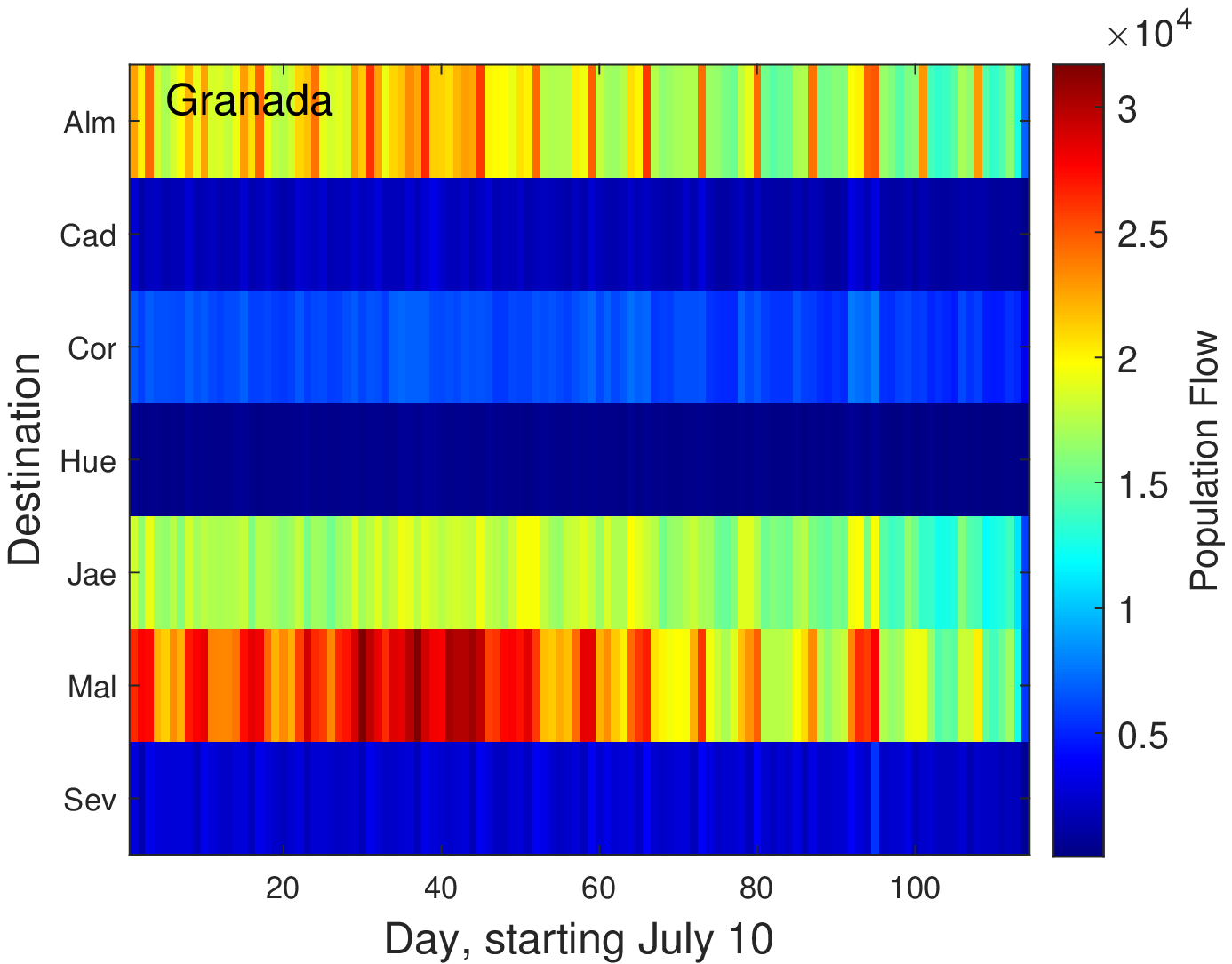} \\
\includegraphics[width=.4\textwidth]{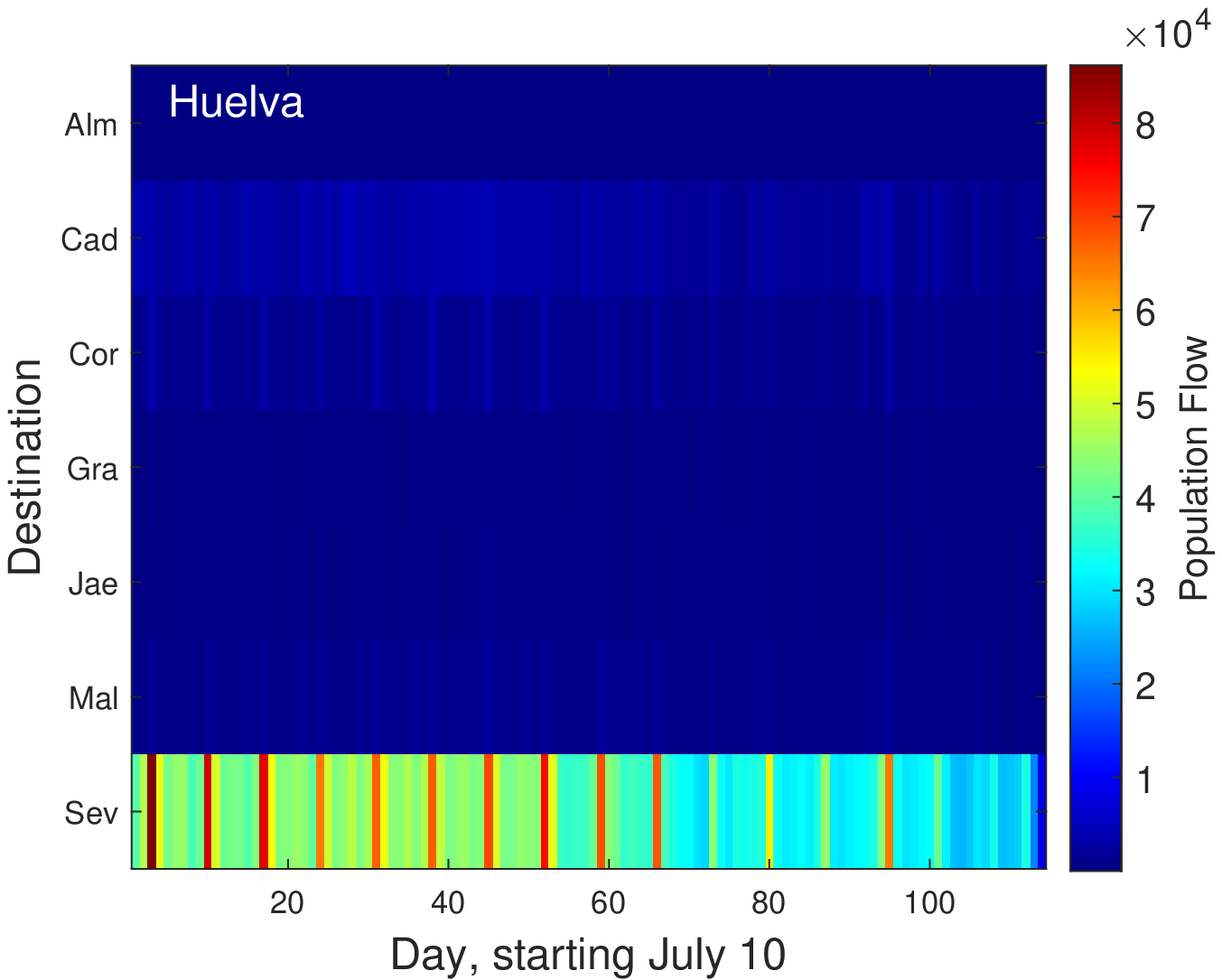} &
\includegraphics[width=.4\textwidth]{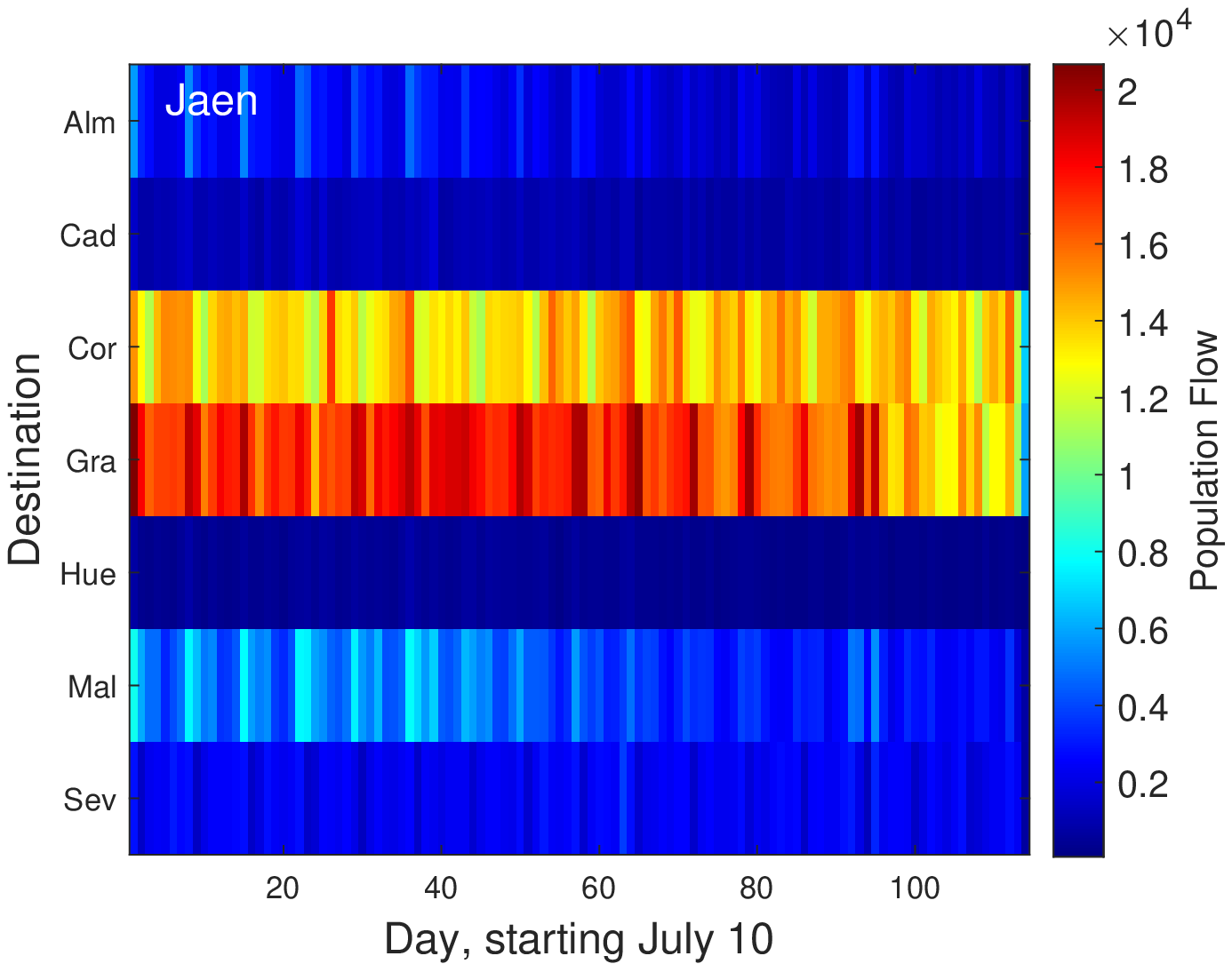} \\
\includegraphics[width=.4\textwidth]{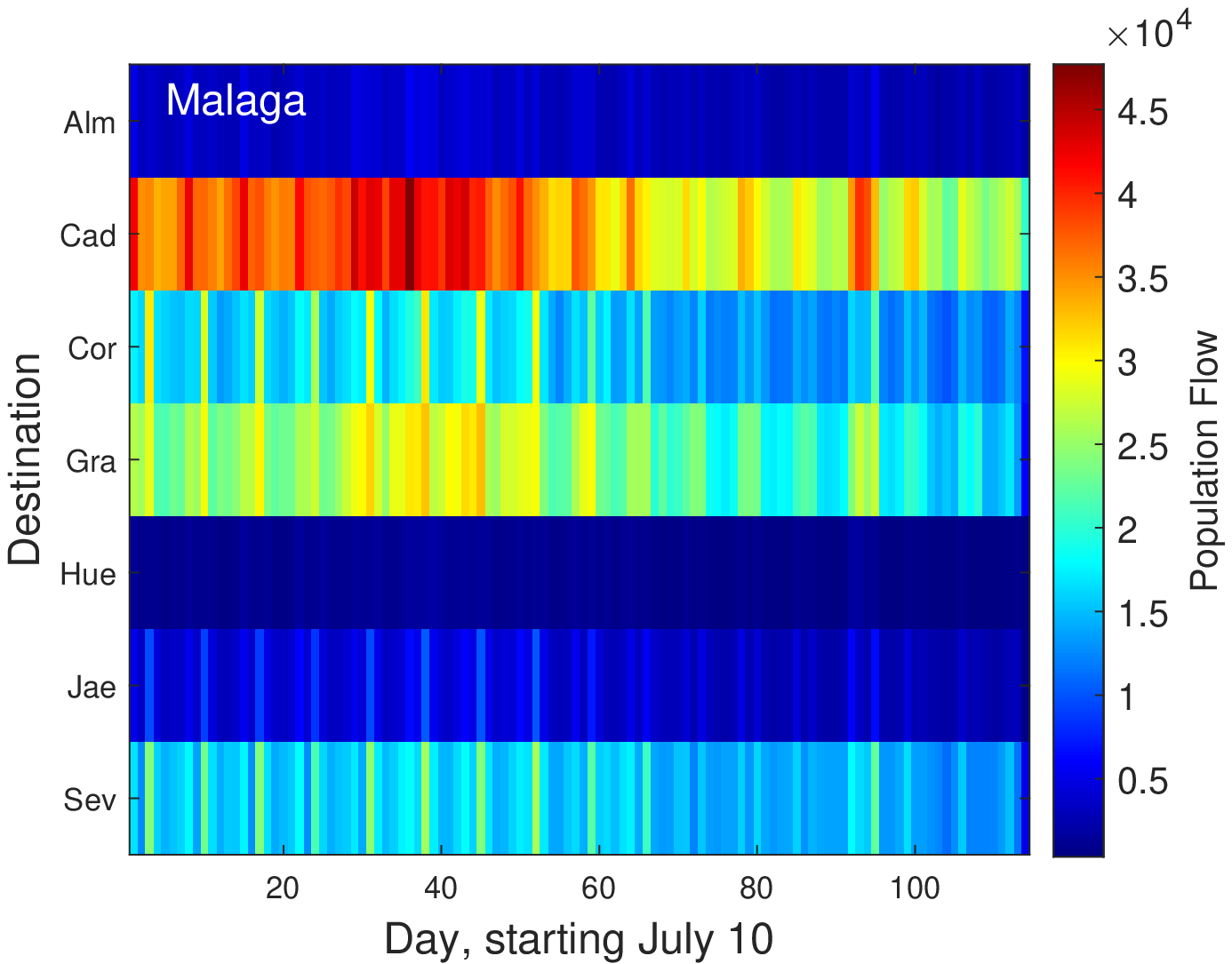} &
\includegraphics[width=.4\textwidth]{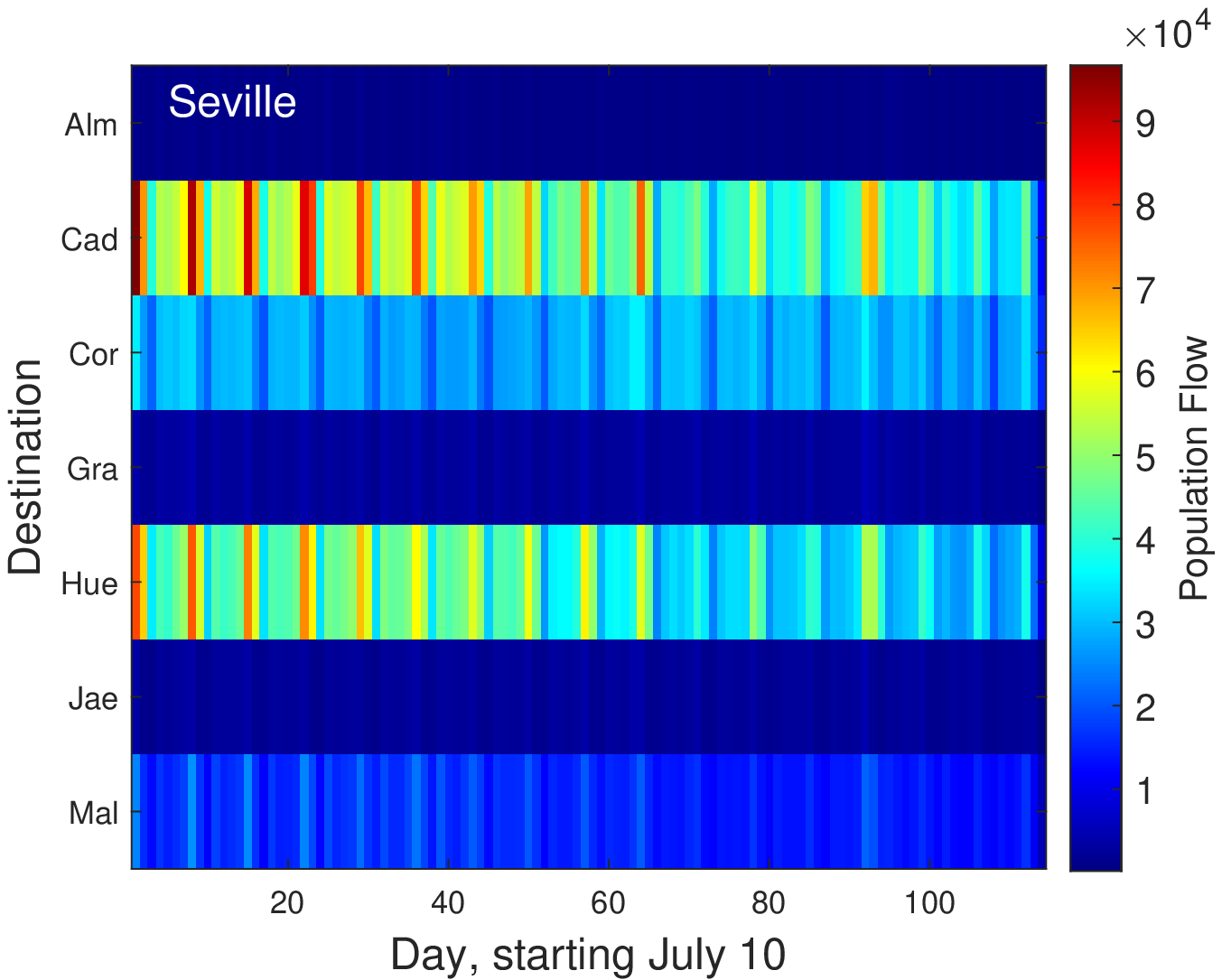} \\
\end{tabular}
}
\end{center}
\caption{Time-varying daily population flows
(number of trips per day, i.e., the daily rate of
traveling $M_{ij}$) for each Andalusian province
as determined from mobile-phone data~\cite{MITMA}. The destinations are shown on the vertical axis.} 
\label{fig:mobility}
\end{figure}

Mobility flows are commonly estimated based on mobile-phone records.
In this work the flows we analyzed are based on a study performed by the Ministry of Transportation, Mobility 
and Urban Agency (Ministerio de Transportes, Movilidad y Agenda Urbana-MITMA)~\cite{MITMA}
that included all of Spain for the period beginning on March 
14, 2020. The main data source was anonymized mobile phone data for more than 13 million mobile lines as well as locations of communication towers and antenna orientations. 
Population data as well as information about the transportation network (airport locations, railways) were leveraged.
Figure~\ref{fig:mobility} shows the time-dependent population flows for each province, i.e., each network node,  of Andalusia
as  determined by the mobile-phone data.  They are shown for the time duration of our study, starting at July 10, 2020 till October 29, 2020 (112 days). Note
the significant time dependence of the inter-province population flows.

Daily mobility data among locations,  such as those provided by mobile phones,   may not be always readily and 
publicly available. When this is the case, other types of data and alternative models 
are used to determine the population flows.
One avenue is to rely on census surveys and base the coupling of the 
epidemiological model on daily commuting data \cite{danon2009, danon2021}. 
In this approach, workdays and weekend, as well as commuters and
non-commuters,  should be distinguished using additional travel surveys. Failure to consider non-work related 
trips may lead to an erroneous slowing down of the epidemic 
\cite{danon2009}. This fine tuning is not required when using time-varying mobility 
matrices as in the present work.

Another avenue is to utilize commonly used trip-distribution modeling techniques, like the 
gravity model to construct the origin-destination (O-D) 
matrix for the metapopulation network. The gravity law is used extensively in the literature to model travel demand
between O-D pairs (e.g., \cite{erlander1990gravity, de2011modelling}). We assume a region where $n$ denotes the
nodes or centroids of the cities in the regional transportation network and $m$ their  highway links. 
A trip matrix element (number of trips per day)
is denoted by $w_{ij}$, where $i$ and $j$ are the origin and destination
nodes of the considered trip, respectively.  Given the population of these cities
and their distances,
%between O-D pairs $ij$, 
the O-D matrix elements are computed by
\begin{align}
    & w_{ij} = C\frac{N_i^\alpha N_j^\gamma}{e^{\beta {\textrm{dis}}_{ij}}}
    \label{eq:gravity}
\end{align}
\noindent where C is a constant, $\textrm{dis}_{ij}$ is the distance between the
O-D pair ($ij$), 
$\alpha$ and $\gamma$ are parameters associated with the populations
$N_i$ and $N_j$ of the pair ($ij$), 
and $\beta$ is associated with this distance. 
Once the elements of the O-D matrix have been estimated, the force of infection
on susceptible hosts $S_j$ in location $j$ in the metapopulation model, 
which reads 
$\beta_{AS} A_j/N_j, ~ \beta_{IS} I_j/N_j$  in Eqs.~(\ref{eq:susm}, \ref{eq:expm})
is modified as \cite{xia2004}
\[
\beta_{AS} \frac{1}{N_j} \Big (A_j + C N_j^{\gamma}  \sum_{i \neq j} \frac{A_i^{\alpha}}{e^{\beta {\textrm{dis}}_{ij}}} \Big ), ~~
\beta_{IS}\frac{1}{N_j} \Big ( I_j + C N_j^{\gamma}  \sum_{i \neq j} \frac{I_i^{\alpha}}{e^{\beta {dis}_{ij}}} \Big ).
\]

\begin{figure}[!ht]
\begin{center}
\begin{tabular}{ccc}
\includegraphics[width=.3\textwidth]{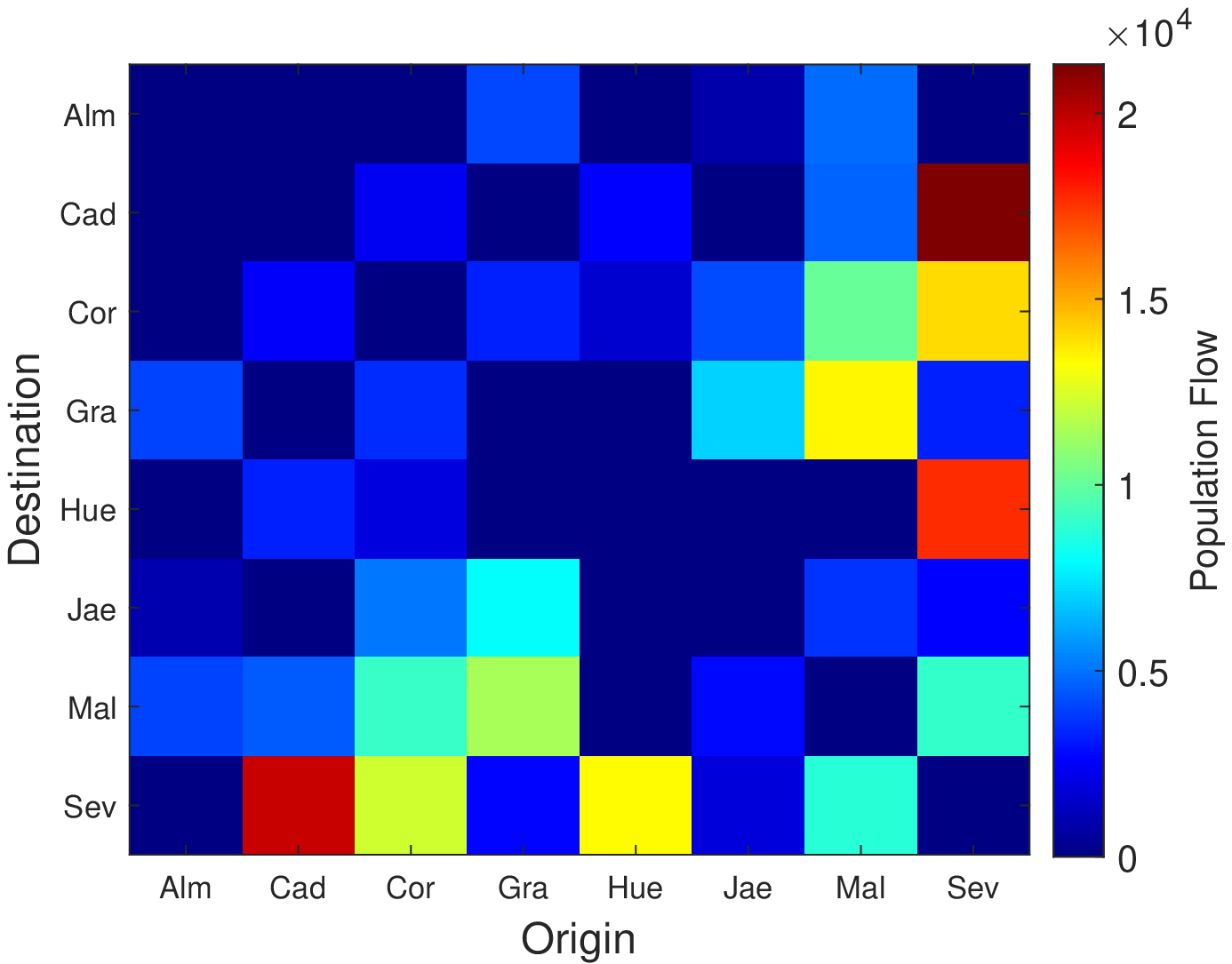} &
\includegraphics[width=.3\textwidth]{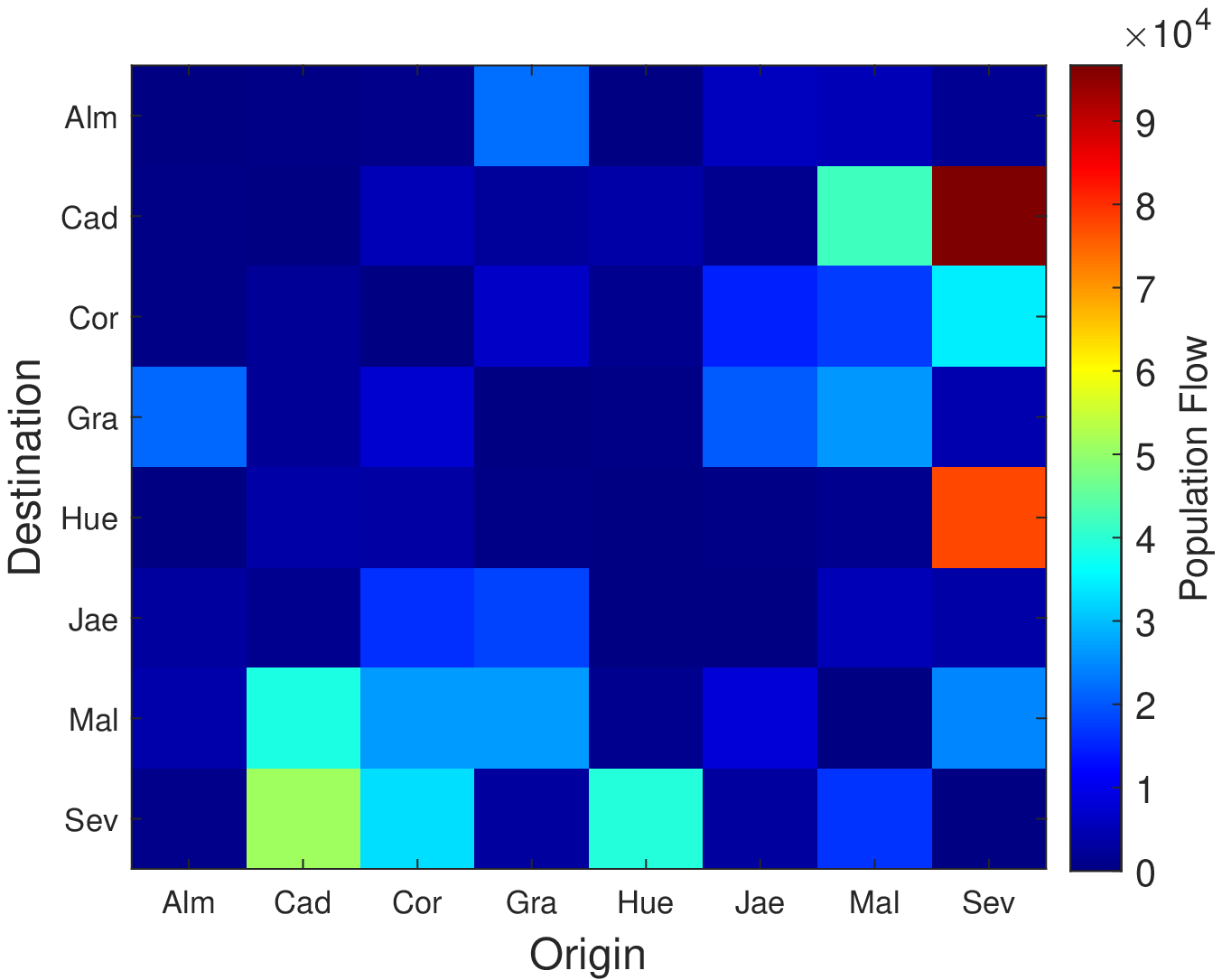} & 
\includegraphics[width=.3\textwidth]{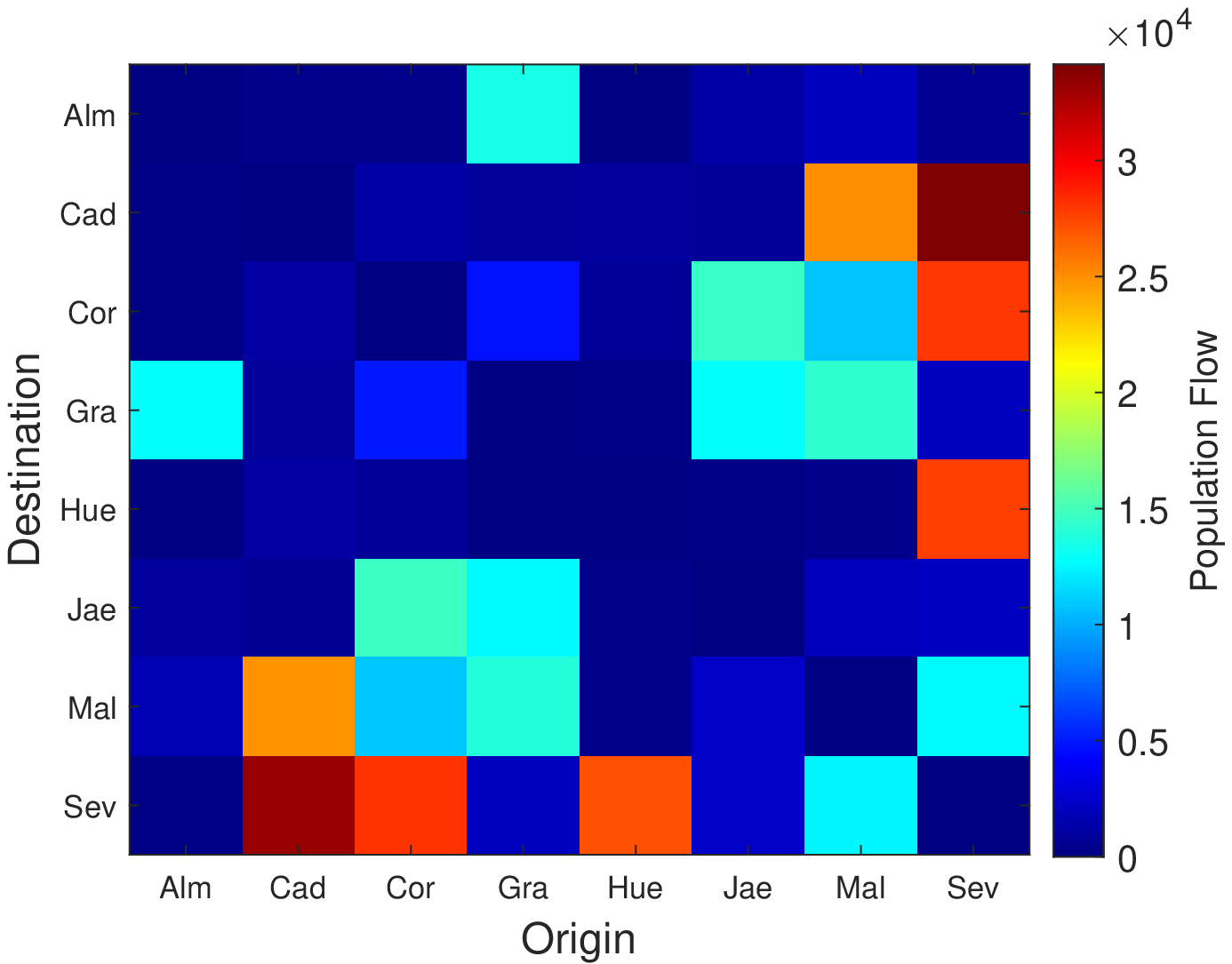}\\
\end{tabular}
\end{center}
\caption{Origin-Destination matrices (trips per day) based 
on the gravity-law method for a 
pre-pandemic day (left panel) and from the
mobility data based on mobile-phone records (people traveling per day) for Friday, July 10, 2020 (middle panel) and Tuesday, 
October 27, 2020 (right panel).}
\label{fig:mobilcompar}
\end{figure}

It is relevant to note that the accuracy of gravity-like models has
received considerable recent criticism~\cite{schlapfer2021,
  simini2012}. In the present work, 
we will not embark on a detailed comparison of a metapopulation
model based on the gravity law and our own approach (based on
time-dependent mobile-phone records). Nevertheless, for completeness,
we would like to illustrate that 
 in the absence of alternative, and possibly quite superior, data
 sets,
 the method can be used to capture some principal features of 
mobility flows in workdays (Friday and Tuesday; no mobility
restrictions in place);
see, in particular, Figure~\ref{fig:mobilcompar} .
More concretely, due  to the scarcity of reported traffic count data --they are averaged over a year -- 
only a static O-D matrix can be obtained. Also, the traffic count data available to us 
were from 2019-2020, namely prior to the pandemic.
The O-D matrices in Figure \ref{fig:mobilcompar} show that the gravity-law 
method roughly 
captures the main mobility trends. For instance, there is substantial
support within the matrix between the rows 2-4 and columns 6-8 (and vice-versa), as well
as e.g. between Seville and Huelva or Malaga etc.
Therefore, in the absence of
more detailed and accurate mobility information, it can be used as an alternative.
The O-D matrix presented in Fig.~\ref{fig:mobilcompar} reproduce the gravity-law
data shown in Table \ref{tab:gravity_od_sp}, Supporting Information.
It should be noted however, that the gravity-law O-D is 
in terms of trips per day, whereas the mobility flows from the Spanish government
are reported  in terms of people traveling per day. To convert one to the other one would  need to know on average the number of people traveling in vehicles.

\section{Results}
\subsection{Period of study and rationale}

The time period considered begins on July 10, 2020 and ends on 
October 29, 2020 (112 days). Since the goal of the present study is to investigate
of the role of mobility on the spread of an epidemic, the period of study was 
chosen to ensure the following. First, that there would be
no imposed mobility restrictions except at the end. In fact, on October 
29, the regional government imposed a curfew at nights and closed the border 
with the rest of Spain and limited the mobility between the provinces.
Second, the period should include the initial exponential growth of 
the epidemic peak. That is the reason why we chose to perform our analysis
up to the end of October 2020, and not longer, as afterwards
the mobility patterns were modified due to the imposed restrictions on travel.

\subsection{Parameter fitting and model predictions}
We first use the model of Eqs.~(\ref{eq:sus})-(\ref{eq:fat}) for the entire region of Andalusia.
We use the norm
\begin{align}
    \mathcal{N} = \sum_{i} \left(\frac{D_{\mathrm{num}}(t_i)}{D_{\mathrm{obs}}(t_i)}-1 \right)^2
    \label{eq:rss}
\end{align}
as the objective function. We minimized it to fit the fatality data $D_{\mathrm{obs}}(t_i)$, where $t_i$ stands for day $i$,
since our start  point of  July 10, 2020, and $D_{\mathrm{num}}(t_i)$, denotes
the fatality estimate for the same day, obtained from the model.
It is worth noting again that we are not attempting to fit to case data, since these
are believed to be significantly less 
reliable than fatality data, due to under-reporting as has
been the case  in other countries as well~\cite{cuevas2021, george2022}.

In addition to the seven parameters shown in (\ref{eq:params}), we also
obtain estimates for the initial parameters $I_0, A_0, E_0$ when the entire autonomous community
of Andalusia is considered.
We performed 500 optimizations with an initial guess for each parameter
uniformly sampled within a pre-specified range. The upper and lower limits of the variation
ranges were used as boundaries in the constrained minimization algorithm
(implemented in Matlab via the \texttt{fmincon} function).
The outcome of the fitting, for values taken from July 10 to October 1, 2020 (84 days), allowed
us to retrieve an approximation for the initial values for the $I$, $E$ and $A$ compartments.
Their median
(they had a very small dispersion) was used as initial condition for the metapopulation model
(\ref{eq:susm}-\ref{eq:totalm}),
weighted by $\omega_j=C_j/C$, with $C_j$ being the number of cases in the $j$-th province in
the period from July 4 to July 10 and $C$ the total number of cases in Andalusia in the whole period.
We minimized the norm
\begin{align}
\label{eq:norm_meta}
    {\mathcal N}=\sum_{j=1}^{8} {\mathcal N_j}
\end{align}
with
\begin{align}
\label{eq:norm_province}
    {\mathcal N_j}=\sum_i\Big{(} \frac{D_{j,\mathrm{num}}(t_i)}{D_{j,\mathrm{obs}}(t_i)}-1\Big{)}^2,
\end{align}
and $D_{j,\mathrm{num}}(t_i)$ and $D_{j,\mathrm{obs}}(t_i)$ being, respectively, the fatality estimate and data for the day $t_i$ at province $j$. In the metapopulation model, we focused on two values of $\theta$, $\theta=1$ and
$\theta=0$, which will be denoted as the metapopulation model with and without mobility,
respectively. As mentioned previously, the coupling matrices $M_{ij}$ were obtained from \cite{MITMA}.
\begin{figure}
    \centering
    \includegraphics[width=0.65\textwidth]{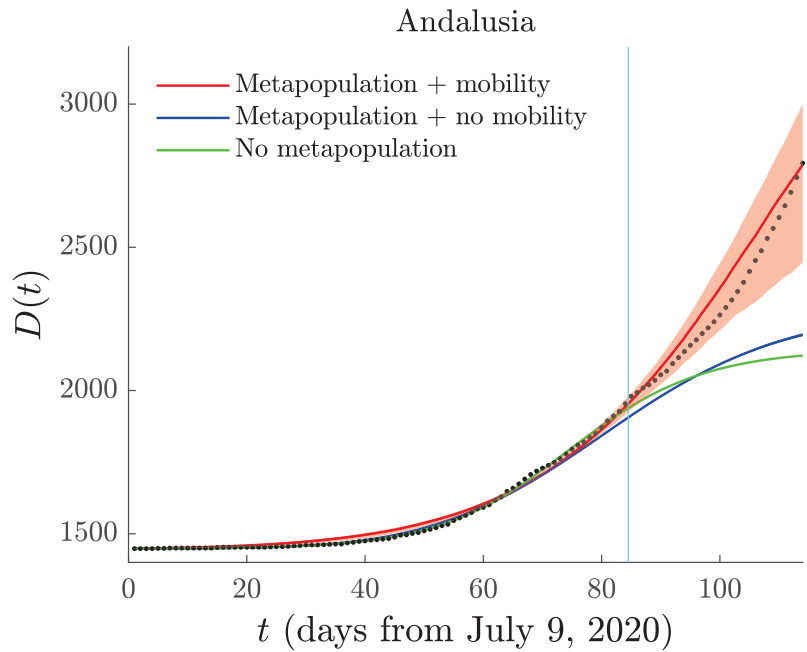}
    \caption{Model fit and prediction of
    the fatalities time series for the entire region of Andalusia.
    Data points are shown as black dots, the output of the metapopulation model with mobility ($\theta=1$) is shown as a red curve, 
    the output of the metapopulation model with mobility turned off ($\theta=0$) is shown as a blue curve, 
    and the fit to the ODE model (Eqs.~\ref{eq:sus}-\ref{eq:fat}) is shown as a green curve. The light blue vertical line
    corresponds to the date when fitting stops (day 84) and prediction begins.
    The interquartile range is highlighted in red.}
    \label{fig:Andalusia}
\end{figure}

Figure \ref{fig:Andalusia} shows the fit of the SEAIR model of Eqs.~(\ref{eq:sus})-(\ref{eq:fat}),
no metapopultion,  together with the metapopulation model
(\ref{eq:susm}-\ref{eq:totalm})
with ($\theta=1$) and without ($\theta=0$) mobility for the case of
only Andalusia. Part of the data (the first 84 days, from July 10 to October 1) is used for parameter fitting, and the remaining is used for prediction (till day 112, from October 2 10 to October 29).
We observe that while all three curves are close to each other
and trail the data  points with a satisfactory level of accuracy
during the fitting period (since we are
fitting them to the data),
they diverge afterwards.
Only the metapopulation model with mobility follows the same trend as the
the data in the prediction interval. One possible reason
is that during summer the fatality curves in all provinces behave similarly, 
i.e., they are quite homogeneous, but later on they follow different
trends, and
they become heterogeneous.
Hence, the overall fatality curve, the one corresponding to the entire
autonomous community of Andalusia, diverges from the homogeneous curve.

Another explanation is that it is possible to fit different models 
to the same data set, but not all models will be able to make accurate
predictions. This is especially true when fitting to epidemic data in the
period before the inflection point of the epidemic peak has been reached \cite{prasse2022}.

\begin{figure}[!ht]
\begin{center}
\resizebox{0.75\textwidth}{!}{\begin{tabular}{cc}
\includegraphics[width=.45\textwidth]{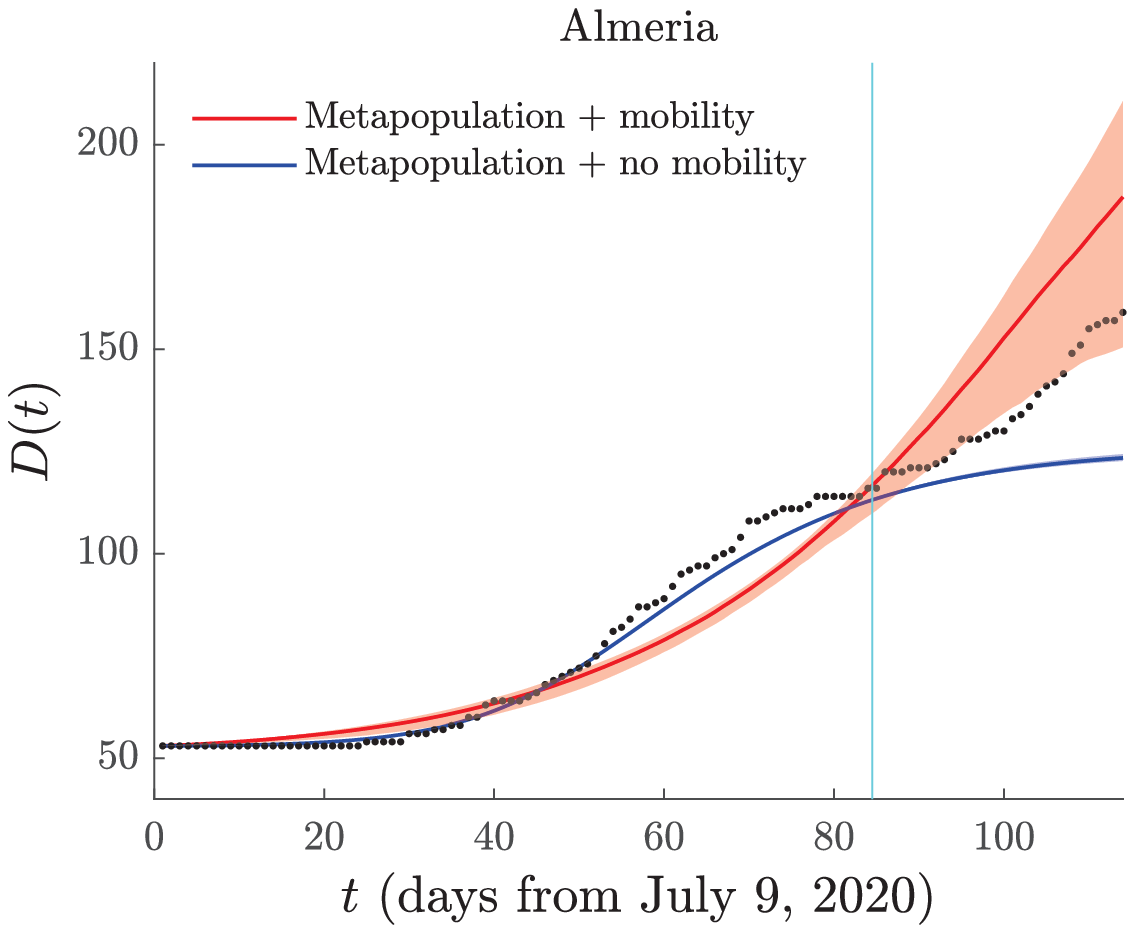} &
\includegraphics[width=.45\textwidth]{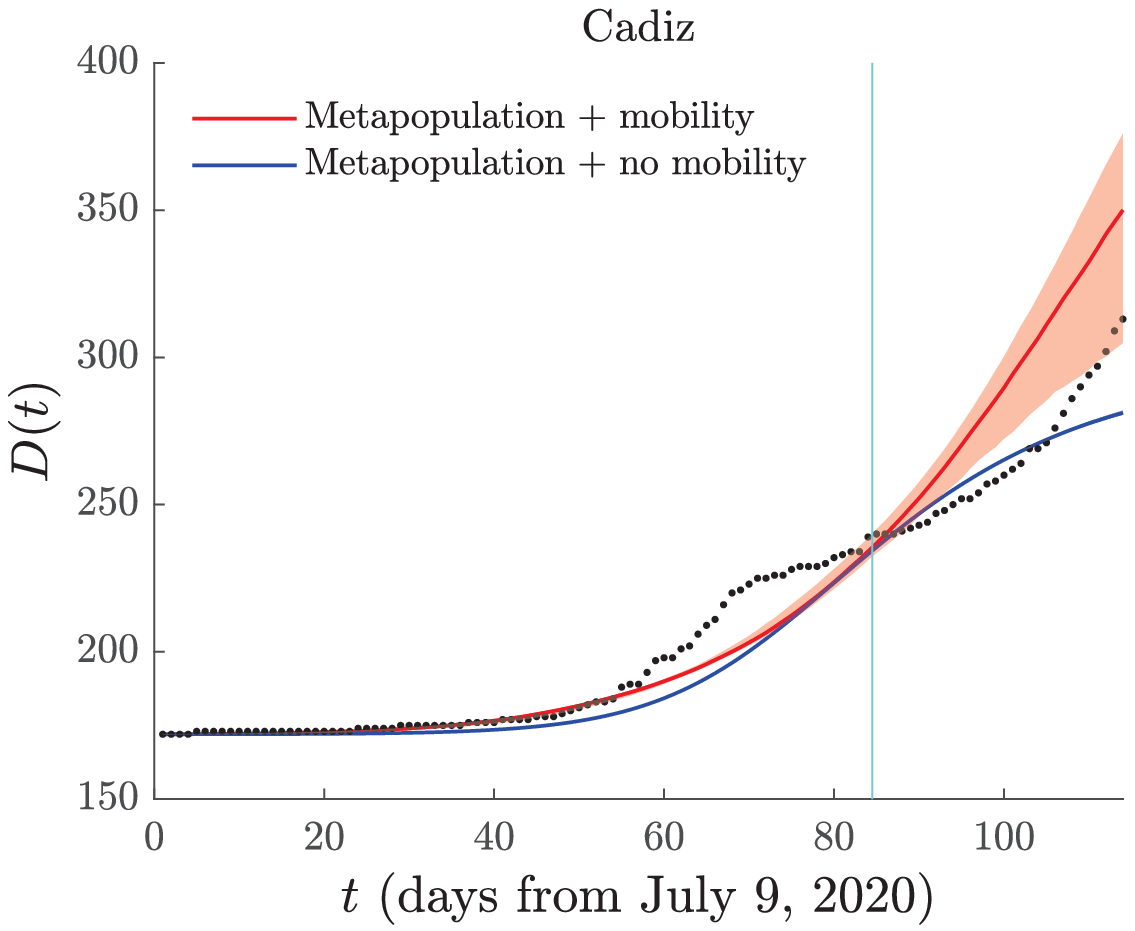} \\
\includegraphics[width=.45\textwidth]{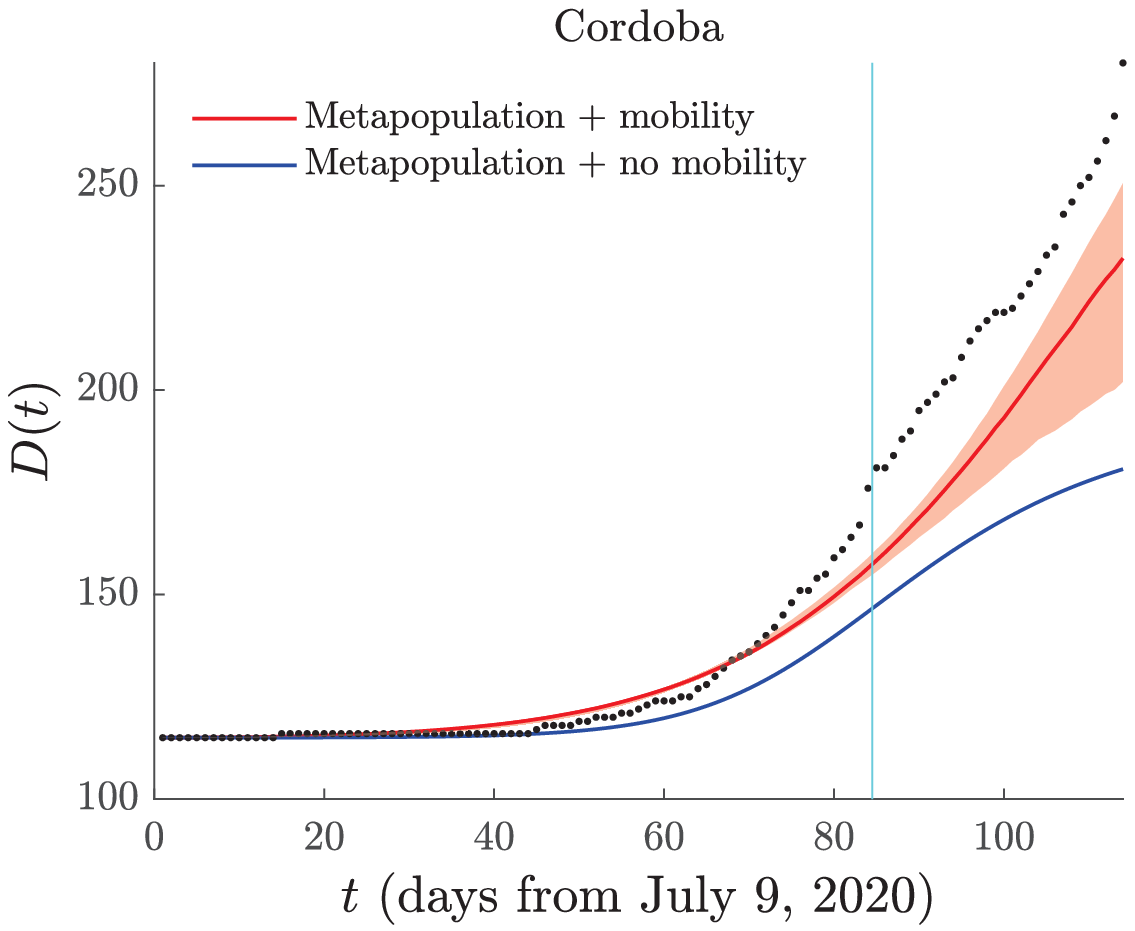} &
\includegraphics[width=.45\textwidth]{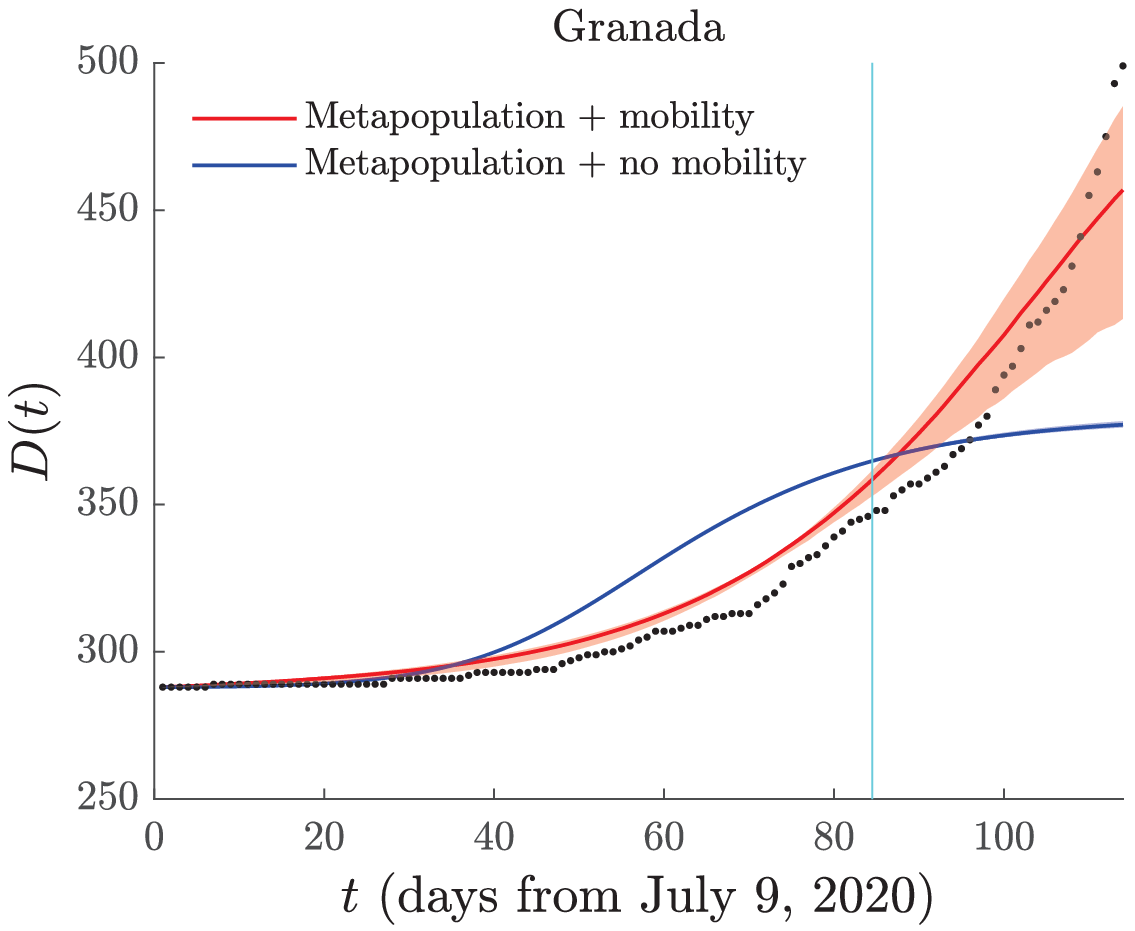} \\
\includegraphics[width=.45\textwidth]{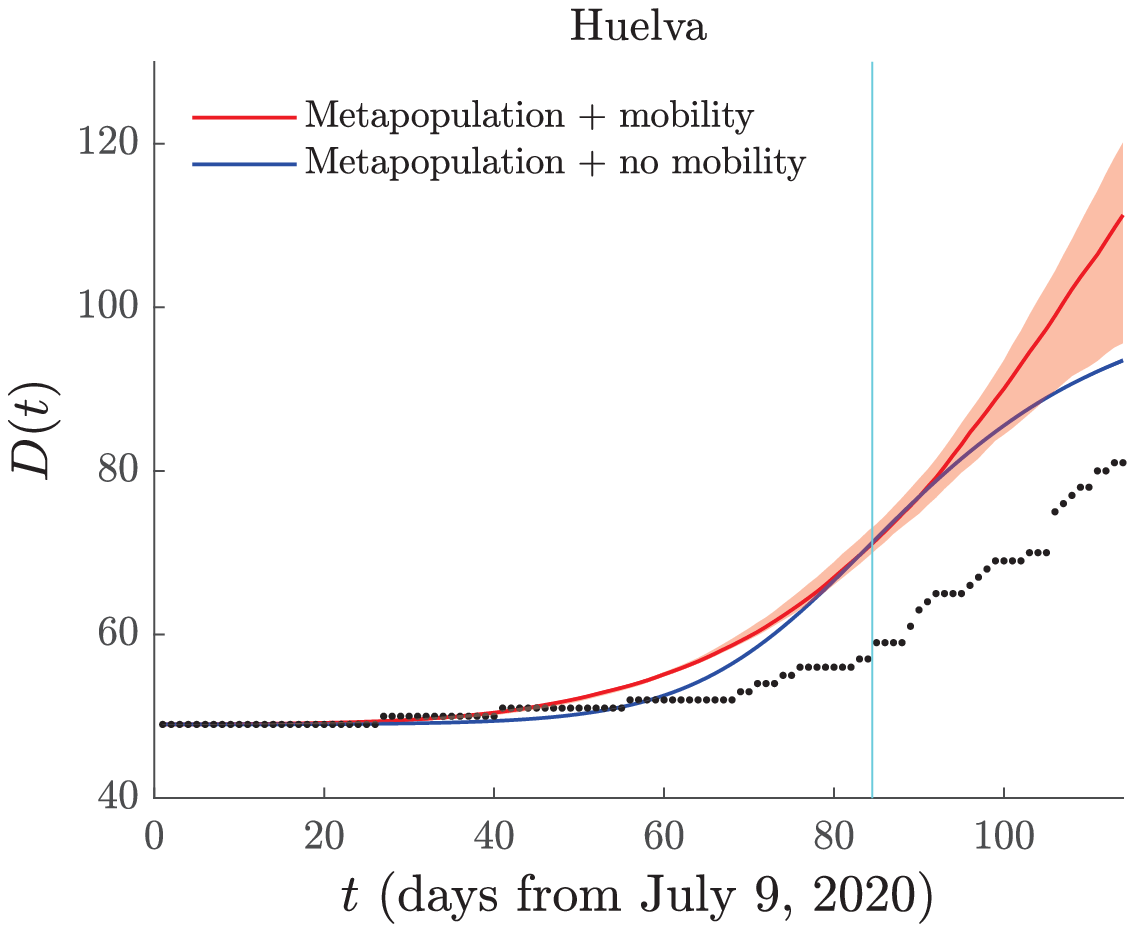} &
\includegraphics[width=.45\textwidth]{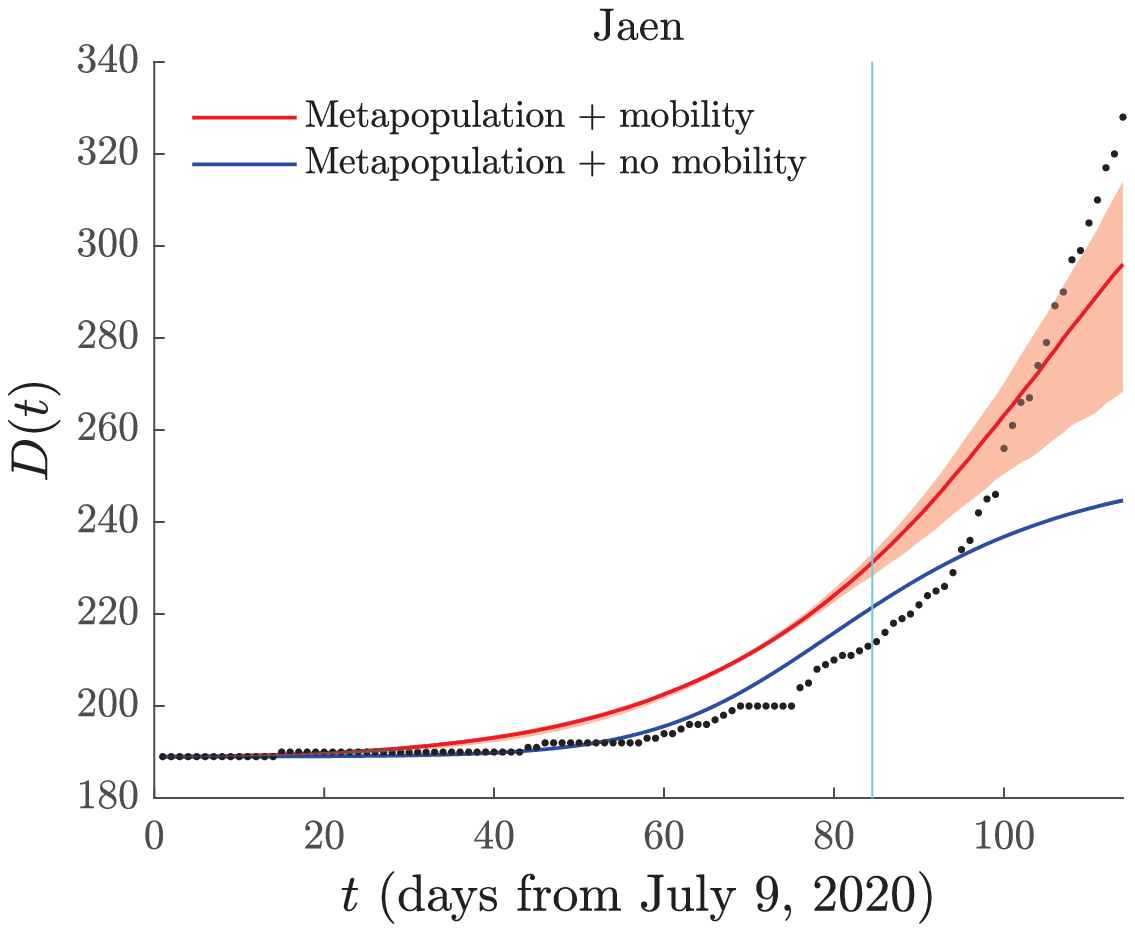} \\
\includegraphics[width=.45\textwidth]{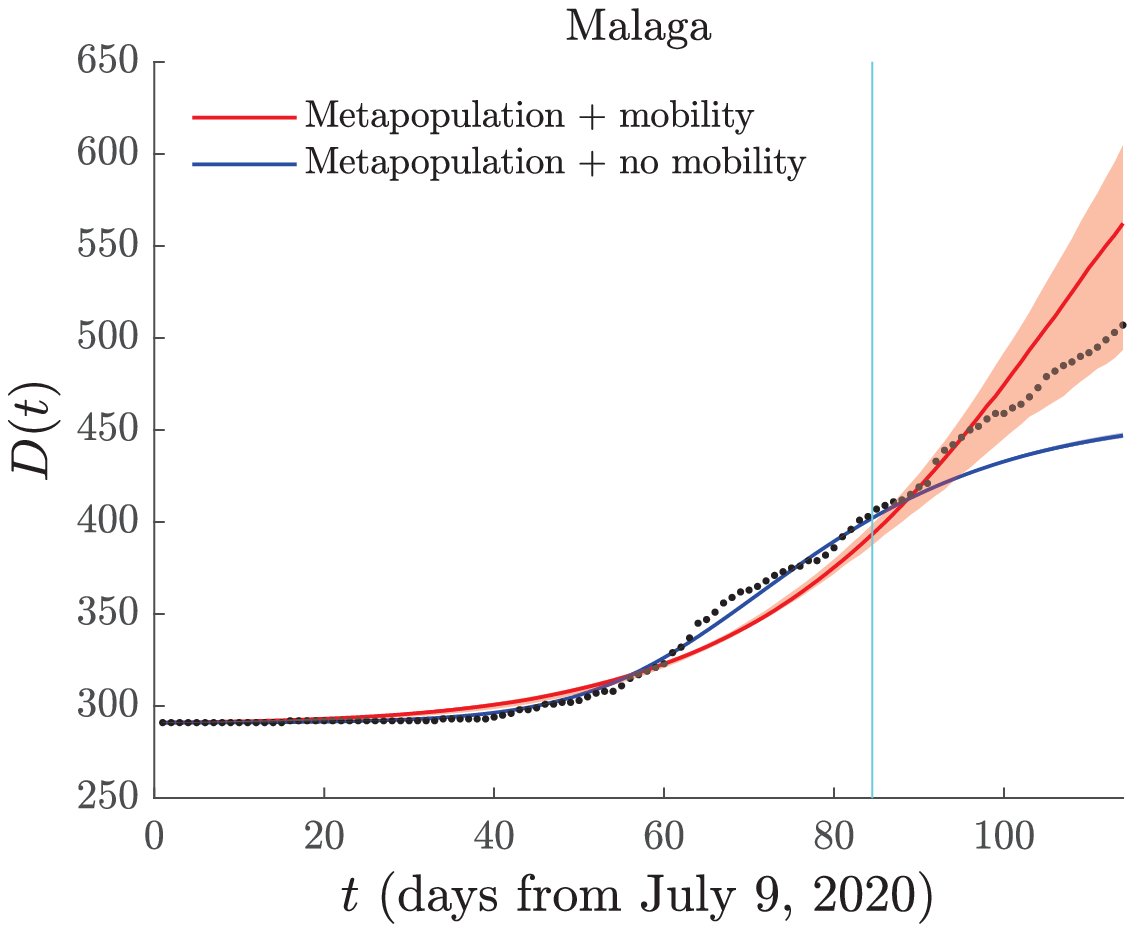} &
\includegraphics[width=.45\textwidth]{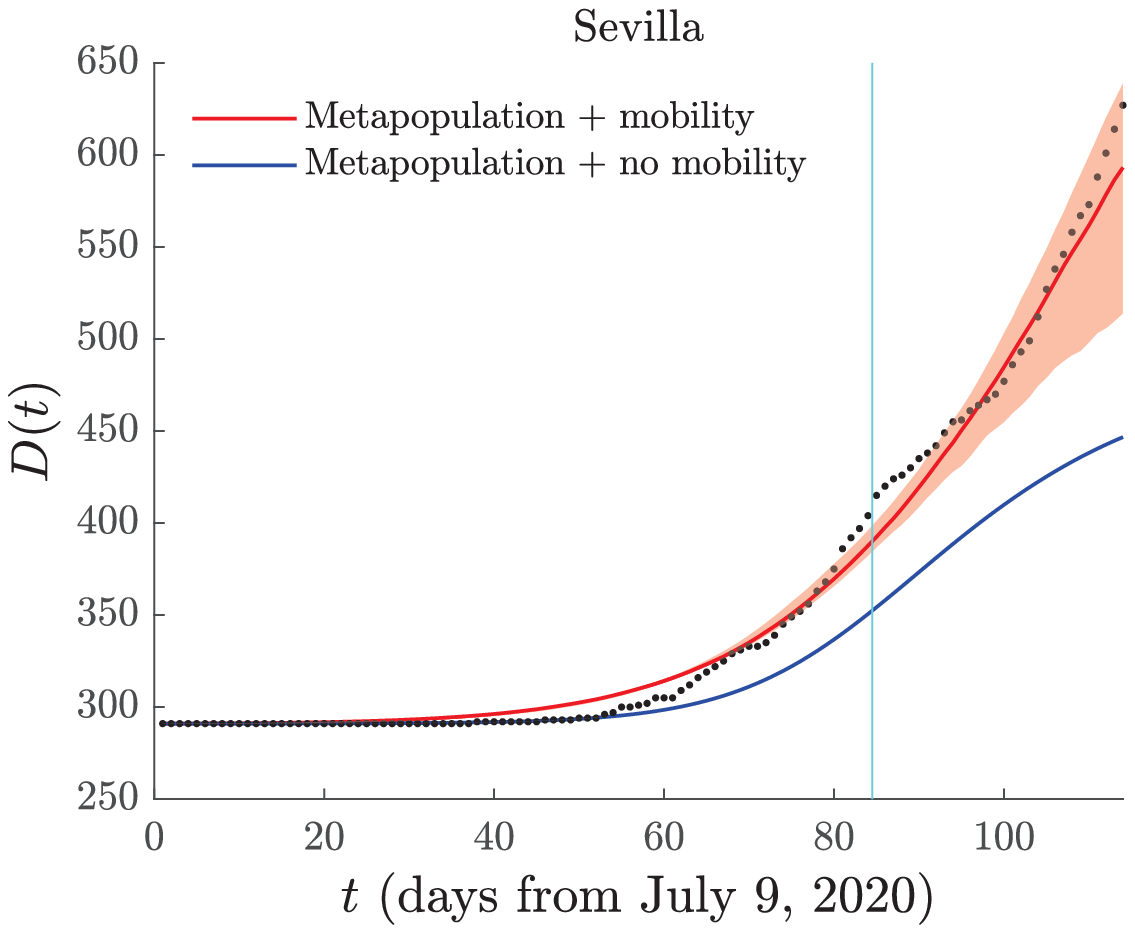} \\
\end{tabular}}
\end{center}
\caption{Fatality data (black dots) and model fit for each province of Andalusia.
    The output of the metapopulation model,
    Eqs.~(\ref{eq:susm})-(\ref{eq:totalm}), with mobility ($\theta=1$)
    is shown as a red curve and the output of the metapopulation model with mobility turned off ($\theta=0$) is shown as a blue curve. 
    The light blue vertical line corresponds to the date when fitting stops (day 84) and prediction begins. The interquartile range is highlighted in light-red for red curve, while for the blue curve, the interquantile range is so narrow that it is not visible in the plot.}
\label{fig:provinces}
\end{figure}

\begin{figure}[!ht]
\begin{center}
\resizebox{0.75\textwidth}{!}{\begin{tabular}{cc}
\includegraphics[width=.45\textwidth]{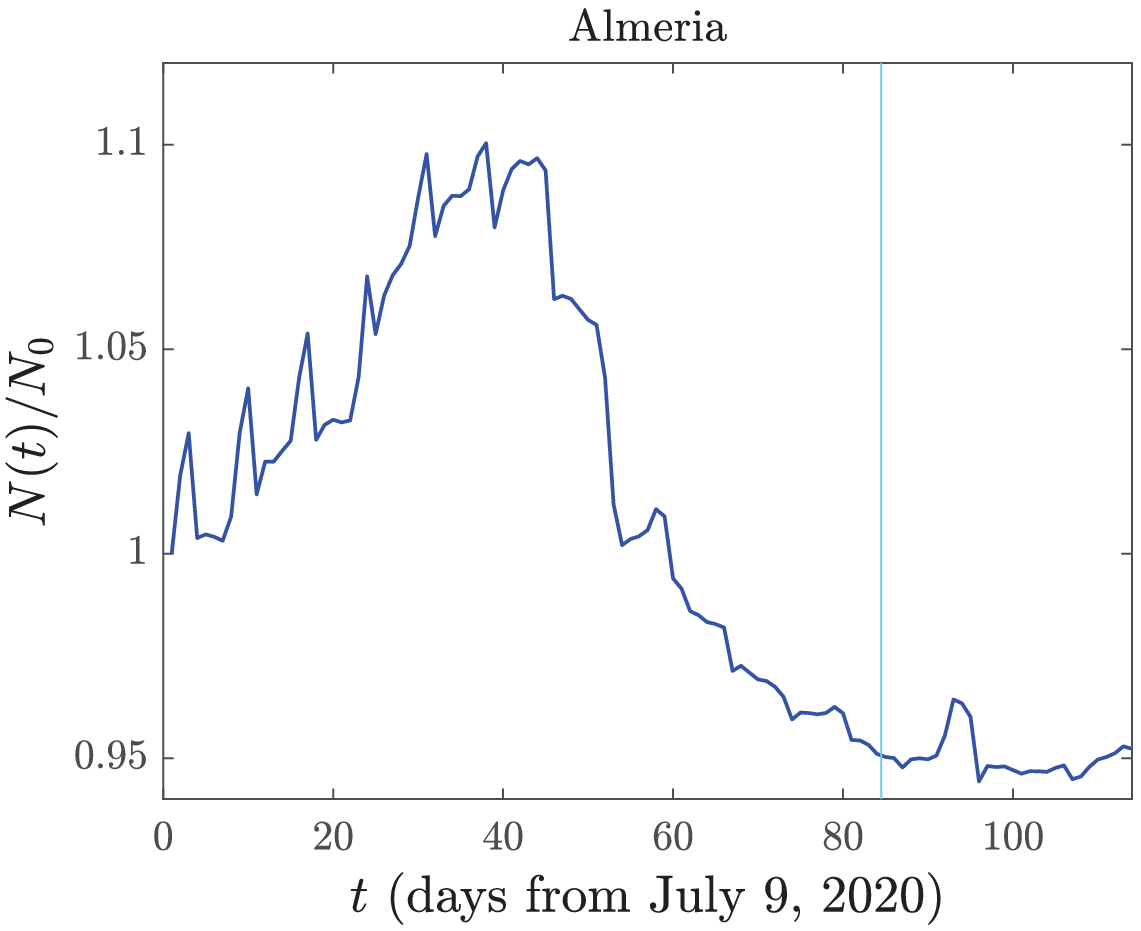} &
\includegraphics[width=.45\textwidth]{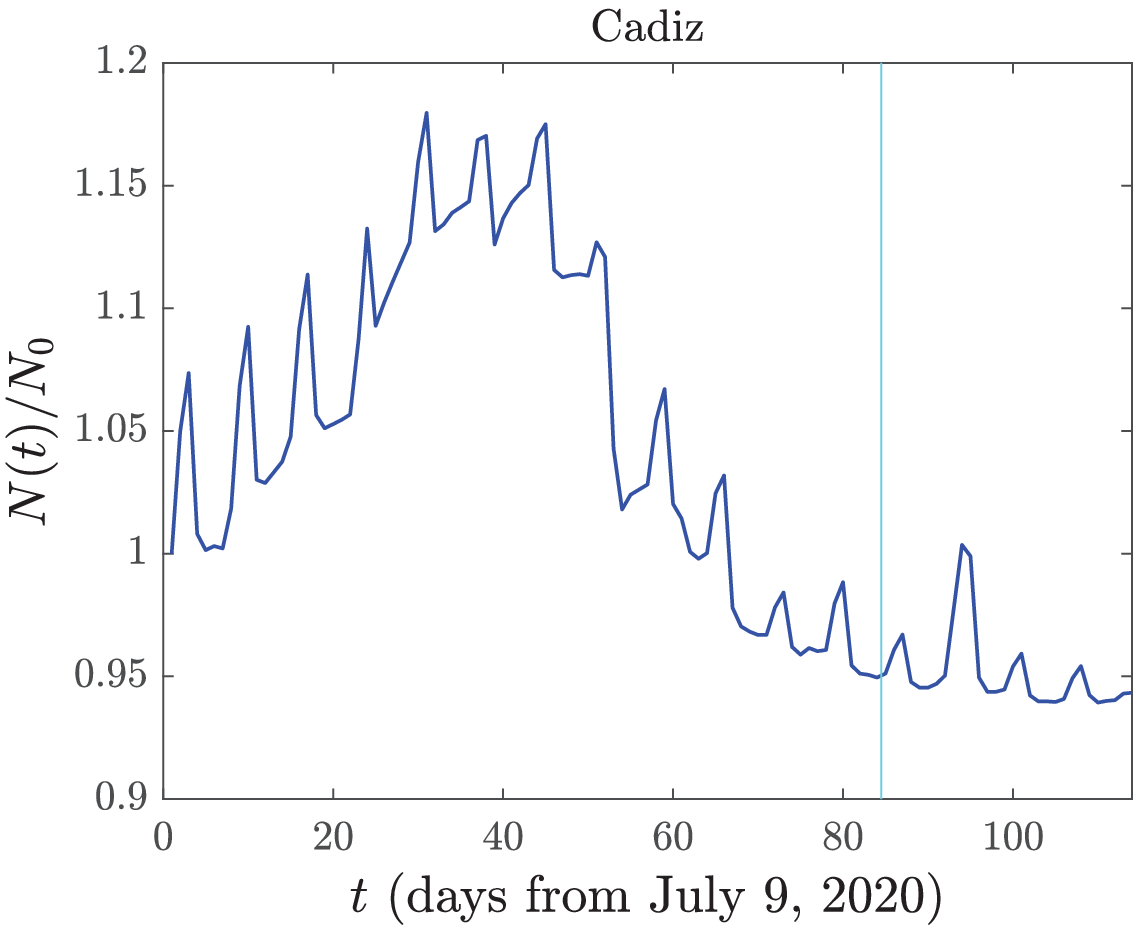} \\
\includegraphics[width=.45\textwidth]{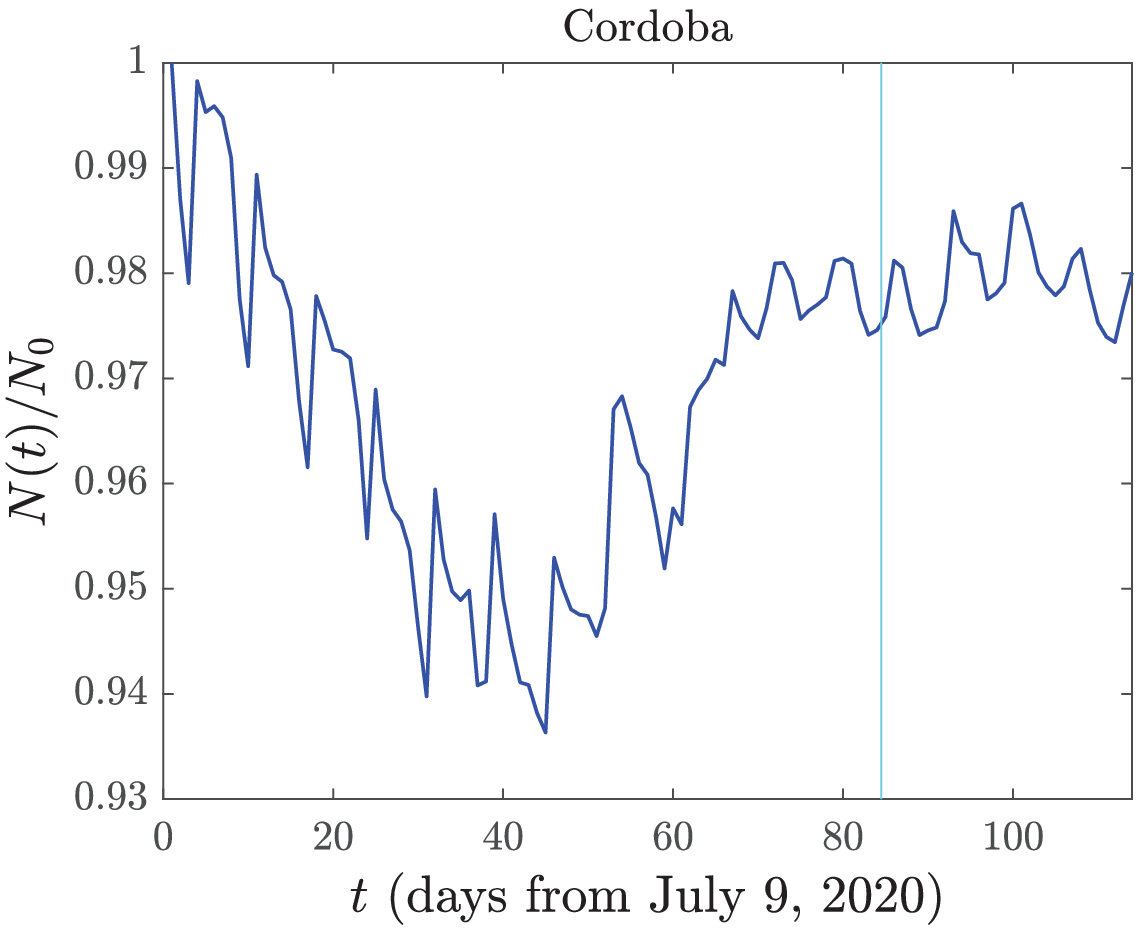} &
\includegraphics[width=.45\textwidth]{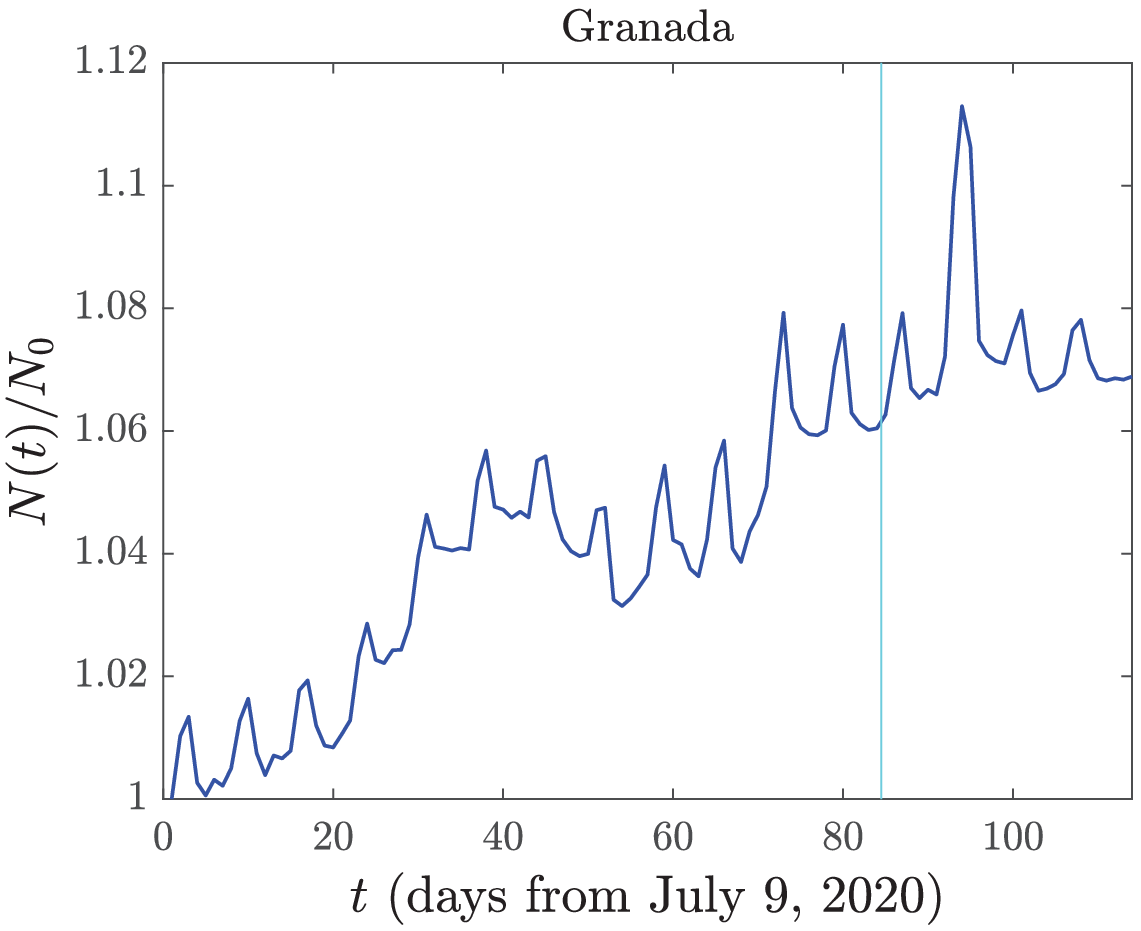} \\
\includegraphics[width=.45\textwidth]{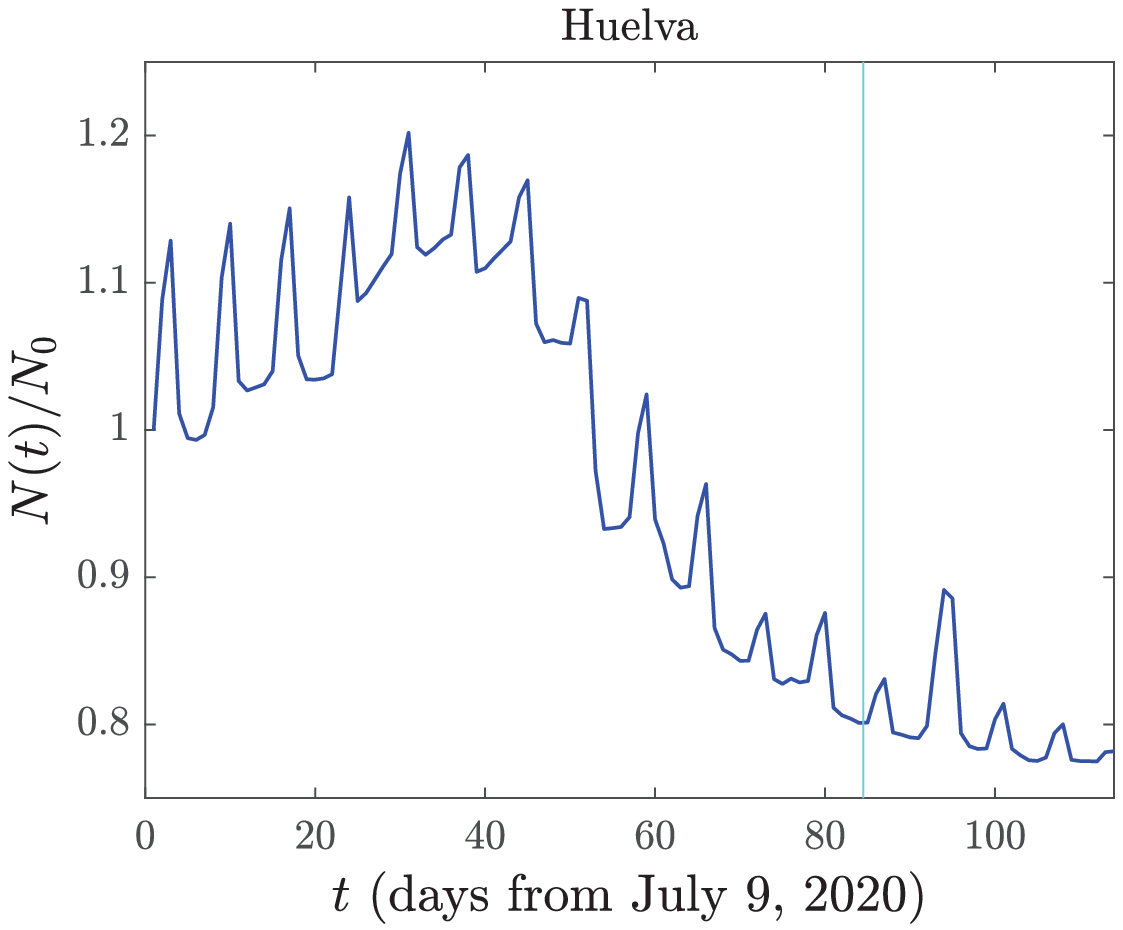} &
\includegraphics[width=.45\textwidth]{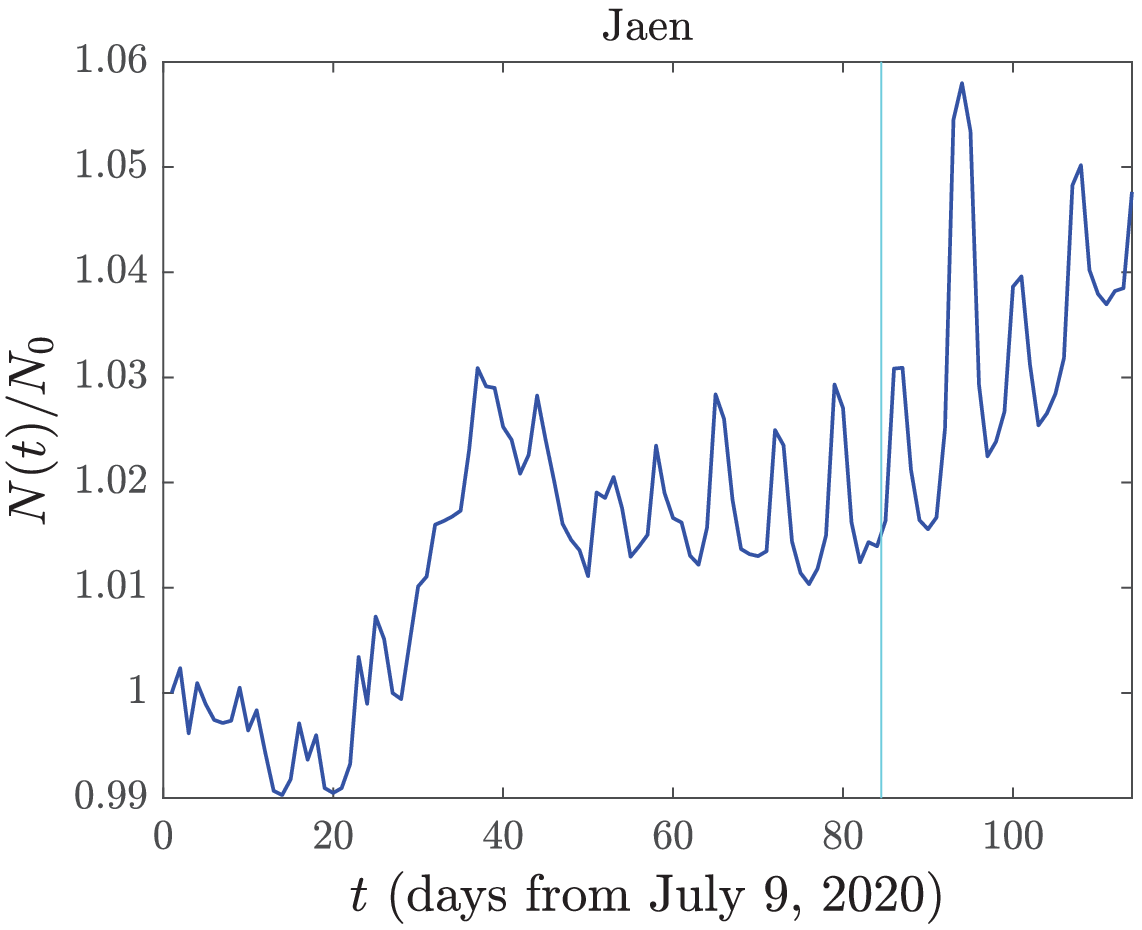} \\
\includegraphics[width=.45\textwidth]{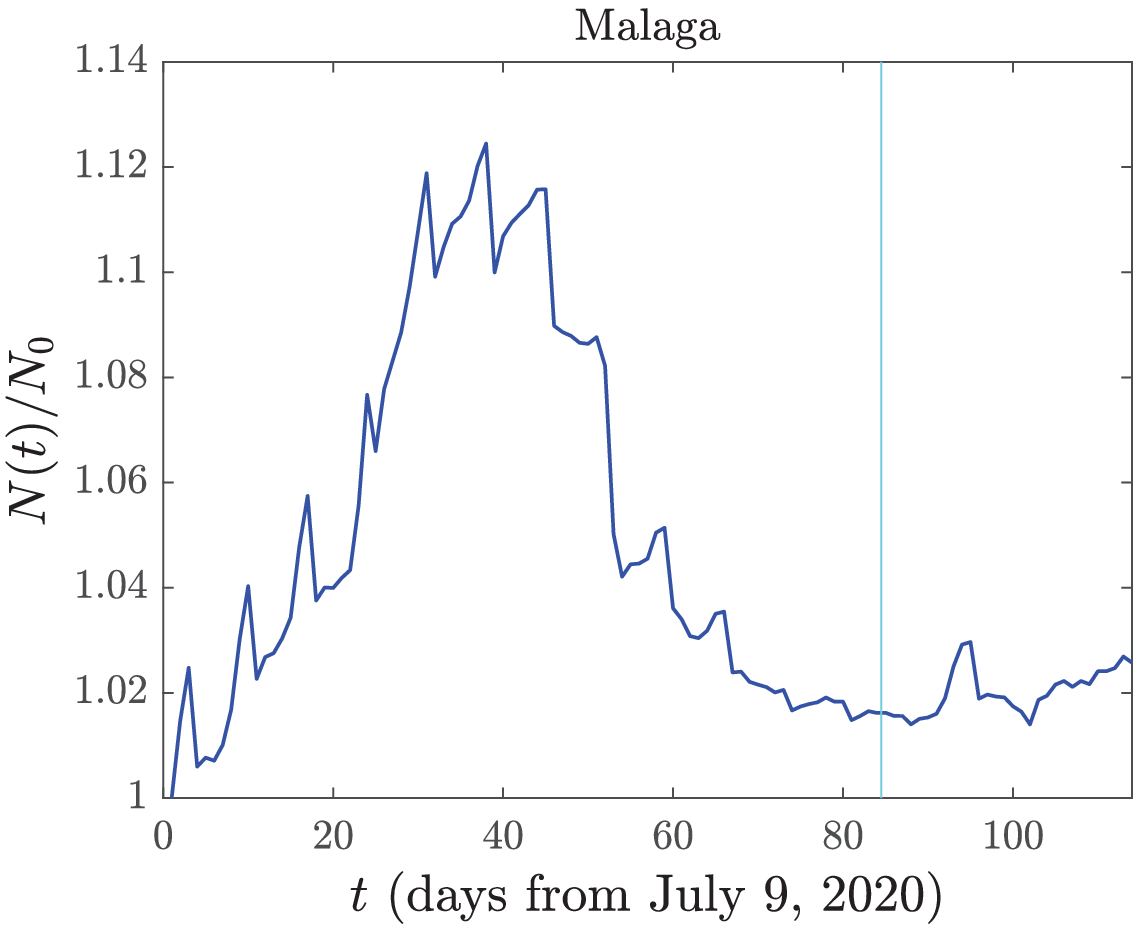} &
\includegraphics[width=.45\textwidth]{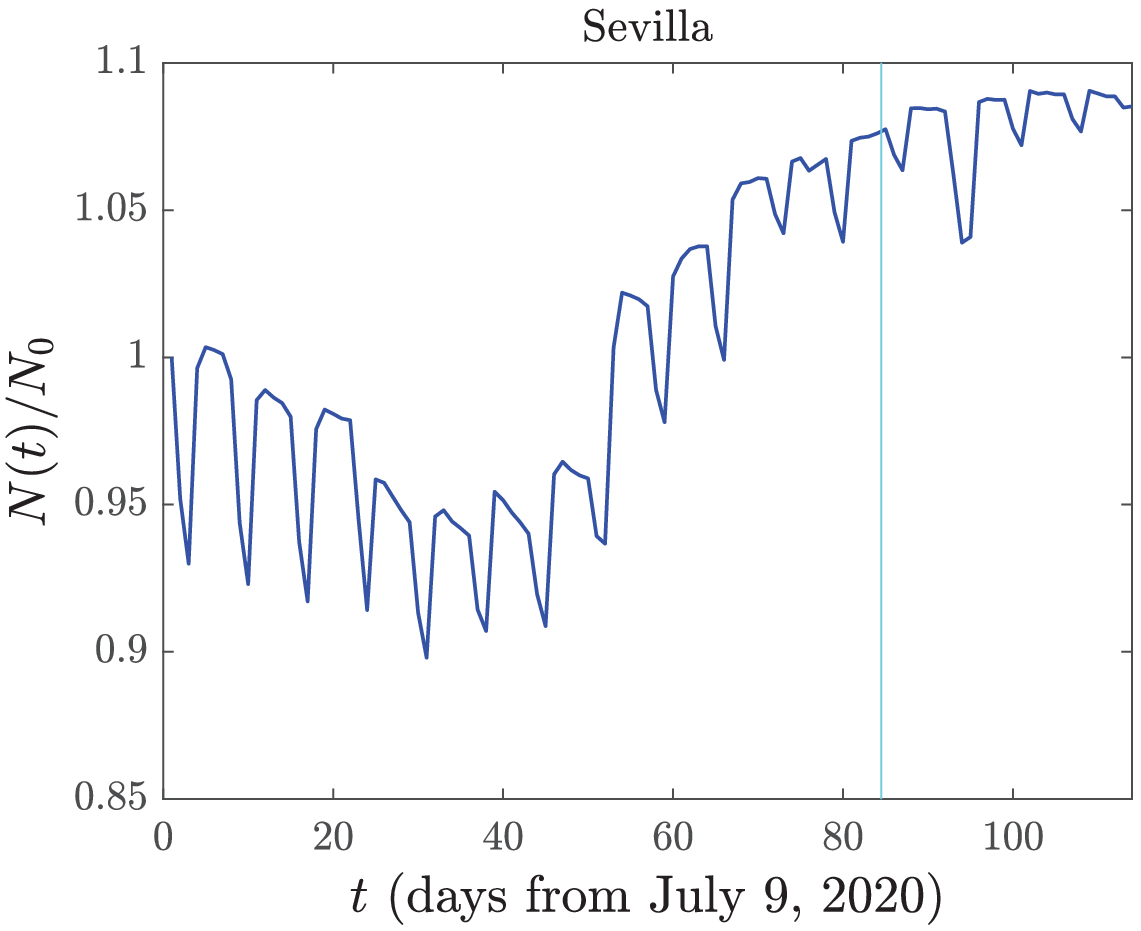} \\
\end{tabular}}
\end{center}
\caption{The ratio of the population of each province $N_i(t)$ to the the initial population $N_0$, the
latter based on data from (\ref{eq:totalm}). The light red vertical line corresponds to the day when fitting stops (day 84)
and validation begins.}
\label{fig:population}
\end{figure}

Further insight on the dynamical evolution of the fatalities in each
of the provinces is provided in Figure \ref{fig:provinces}. 
%In most European countries, due to stringent mitigation measures during 
%the first wave of the pandemic, mobility was limited and behavior was relatively %homogeneous. On the other hand, 
During the months considered in the present study, 
due to relaxation or complete absence of mitigation measures, different nodes of 
the network exhibit different characteristics. This can be attributed to
some nodes being touristic destinations (Malaga, Huelva), others being close to
country borders (Cadiz with Gibraltar and Huelva with Portugal), while yet others undergoing annual exodus over the
summer months (Seville). This is evident in Figures \ref{fig:mobility} and 
\ref{fig:population}, where we show the variation in mobility flows and 
population, respectively, for the eight
provinces forming our network. 

It is relevant to make the following observations in connection with the results.
With the exception of Cordoba (featuring a systematic underestimation within 
the prediction interval  for which
we do not have a definitive explanation)
and Huelva (with a corresponding overestimation 
in the prediction interval), data points typically follow qualitative trends consonant with the interquartile range. 
Huelva is a major vacation hub, both in the summer period and during weekends, mainly from residents in Seville
(who also commonly spend their holidays in the province of Cadiz). 
If people return to their permanent residence to receive treatment and quarantine (or are anyway logged as cases within these regions), 
this may explain the disparity between the observed and predicted fatalities. 
It is worthwhile to note an apparently similar overestimation trend 
within the prediction interval for Cadiz, however, in this case, the 
situation is somewhat less clear, due to an opposite trend within the fitting interval. 
Also, high population density over the summer could partially explain the overestimation: the model is trained with more people residing 
there, who subsequently depart to return to their regular residence locations. 
Also, compared to other provinces, fatalities are relatively small in number, 
which makes it prone to stochastic effects~\cite{Calleri2021,ando21}, as is also evident in the trends of 
the data.

Figure \ref{fig:population} shows the population in each province
during the period of  our study. Two major trends emerge. First, there is a weekly oscillation,
due to increased mobility during the weekends. 
This is due to residents traveling from their primary residences 
to vacation destinations, such as the ones we described 
before between Seville and Huelva or Cadiz; similar patterns are found between other pairwise
transitions: e.g., in the case of Cordoba, such movements happen to and from
Malaga, Seville and Jaen. In any event, the real-time data used in this work provide a clear
picture of the dynamics across the network and the key interactions across
its nodes. Second, there is a
significant variation in the population of most provinces, ranging from mild
(0.99-1.06 in Jaen, 0.93-1 in Cordoba) to extreme (0.75-1.2 in Huelva).
Others, experience a peak in late summer (Almeria, Cadiz,  Malaga) before
their population drops again in October. Granada and Seville exhibit
a reverse behavior, where their population increases in the fall, when
people resume living in their permanent residences. 
This is the seasonal trend that is superposed to the weekly trend. 
A similar observation may be made by considering
Figure~\ref{fig:mobility} where the time-dependent
population flows between any two provinces are shown. 
In line with our above
observations, some clear signatures are obvious,
such as weekly periodicity, overall increased mobility in the  summer months
and other trends, such as the consistent mobility between 
specific pairs of provinces, as discussed above.

\begin{figure}[!ht]
\begin{center}
\includegraphics[width=.95\textwidth]{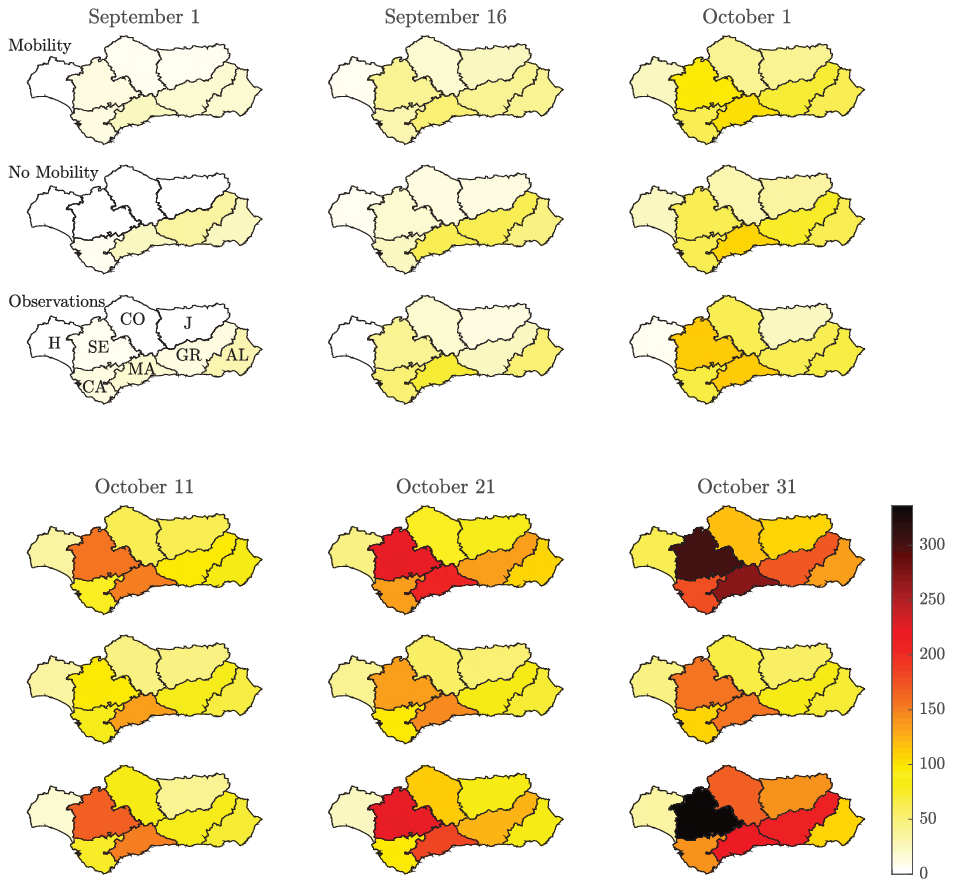}
\end{center}
\caption{Snapshots of the evolution of the number of deaths occurring from (and including) July 10 at each province at different days from September to October. Top and middle row maps correspond to the numerical fit/prediction of the metapopulation model with and without mobility, whereas bottom row maps represent the observed number of deaths. Bottom map in the first snapshot includes the code for the name of each province (AL: Almeria, CA: Cadiz, CO: Cordoba, GR: Granada, H: Huelva, J: Jaen, MA: Malaga, SE: Seville.}
\label{fig:maps}
\end{figure}

Figure \ref{fig:maps} depicts the time evolution of the fatalities
in the form of a heat map. We can observe how the model predicts 
the spreading of the epidemic from Almeria to neighbouring provinces. Note, however,
that in the reported data there was a spot in Malaga, probably caused by people traveling from other places in the
world (a process which is not included in the current work). 
We also observe that Seville and Malaga are the provinces
that eventually exhibit the highest number of fatalities,
an observation correlated to their higher population. 
The maps also show that the model without mobility predicts a very much smaller number of fatalities
than the model with mobility. Although at an early stage of the
prediction  all models are fairly comparable,  later on within the prediction interval
the model with mobility is  significantly more accurate towards
predicting the spread of the epidemic within the metapopulation network
than the model without it. 
Both the detailed (individual province,  cf. Fig.~\ref{fig:provinces}) quantitative findings, and this overarching figure
are convincing, in our view, of the relevance at such regional levels of the 
consideration of metapopulation approaches. Additionally, the concrete trends
that our mobility data reveal illustrate the relevance of the dynamic  
consideration of the coupling matrices $M_{ij}$.

The best-fit parameters are shown in Table \ref{tab:tab1}, whereas
Table \ref{tab:tab3} presents the initial conditions for each province in the
metapopulation model. For the initial condition of the population ($N_0$) we took the
census data for January 1, 2020
\cite{popdata}.
We must note that the data we compared to
correspond to the day where events actually occurred
and were extracted from the 
data available in the National Epidemiological Center of Spain
\cite{CNE}, as well as that each fatality is assigned to the residence province.

\begin{table}[!htb]
\centering
\caption{Best fit parameter values}
\resizebox{\columnwidth}{!}{
\begin{tabular}{cccc}
\toprule
Parameter & Median (interquartile range) & Median (interquartile range) & Median (interquartile range) \\& (No metapopulation) & (Metapopulation w/o mobility) & (Metapopulation w/ mobility)\\
\hline
$\beta_{IS}$ & 0.475749 (0.439256--0.497102) & 0.428599 (0.422961--0.446900) & 0.394443 (0.274636--0.511767)\\ 
$\beta_{AS}$ & 0.483056 (0.469496--0.506530) & 0.490886 (0.469169--0.492347) & 0.412055 (0.287619--0.513972)\\ 
$\kappa_A$ & 0.338381 (0.333010--0.347806) & 0.493038 (0.469970--0.494282) & 0.313193 (0.254518--0.372569)\\ 
$\kappa_I$ & 0.280513 (0.270371--0.287488) & 0.426593 (0.377388--0.453029) & 0.278822 (0.227327--0.331485)\\ 
$\gamma_A$ & 0.317751 (0.308228--0.324681) & 0.461193 (0.421694--0.474062) & 0.318094 (0.262533--0.384494)\\ 
$\kappa_R$ & 0.334552 (0.326885--0.345297) & 0.192424 (0.191560--0.201661) & 0.320300 (0.242333--0.417852)\\ 
$\kappa_D$ & 0.000106 (0.000102--0.000113) & 0.000056 (0.000054--0.000060) & 0.000455 (0.000287--0.000558)\\ 
$E_0/I_0$ & 1.59 (1.57--1.61) & --- & --- \\
$A_0/I_0$ & 1.60 (1.58--1.63) & --- & --- \\
$I_0$ & 587.05 (582.71--591.87) & --- & --- \\
\bottomrule
\end{tabular}}
\label{tab:tab1}
\end{table}

\begin{table}[!htb]
\centering
\caption{Initial conditions}
\begin{tabular}{ccccc}
\toprule
Province & $I_0$ & $E_0$ & $A_0$ & $N_0=S_0$\\
\hline
Almeria & 175 & 278 & 281 & 727945 \\
Cadiz & 26 & 41 & 42 & 1244049 \\
Cordoba & 9 & 15 & 15 & 781451 \\
Granada & 246 & 391 & 395 & 919168 \\
Huelva & 7 & 11 & 11 & 524278 \\
Jaen & 14 & 23 & 23 & 631381 \\
Malaga & 95 & 150 & 152 & 1685920 \\
Seville & 14 & 23 & 23 & 1950219\\
\bottomrule
\end{tabular}
\label{tab:tab3}
\end{table}

\subsection{Effective reproduction number}
\label{sec:R0}

Given the importance of the reproduction number during the initial stages of an epidemic wave, we use 
the Next Generation Matrix approach \cite{diekmann1990} to evaluate 
the effective reproductive number $R_t$.  In doing so, we treat
this epidemic wave as a "new epidemic" assuming  that most of the population is still susceptible. {Hence, the calculated effective
reproduction number refers to the first day of our simulations, July 10th, 2020.}
The reproduction number for the one-node model 
(\ref{eq:sus}-\ref{eq:fat}) (with either $N-D$ in
the denominator or the simplified version $N-D \approx N$) is
\begin{align}
R_t  = \frac{\kappa_A}{\kappa_A+\kappa_I} \frac{\beta_A}{\gamma_A} +
\frac{\kappa_I}{\kappa_A+\kappa_I} \frac{\beta_I}{\kappa_D+\kappa_R}.
\label{eq:R01node}
\end{align}
The first term is the contribution to $R_t$ from asymptomatic hosts $A$ while 
the second is the contribution from the symptomatically infectious hosts $I$.
Each term represents the fraction of asymptomatic $\kappa_A/(\kappa_A+\kappa_I)$
or symptomatically infected  $\kappa_I/(\kappa_A+\kappa_I)$
hosts generated in the lifespan of an
exposed host $E$,  
or equivalently the  fraction of individuals reaching $A$ or $I$ after 
going through state $E$, multiplied by the number of new infected 
hosts generated in the lifespan of the corresponding infectious host,
$\beta_A/\gamma_A$,  $\beta_I/(\kappa_D+\kappa_R)$, respectively.

Using the estimated parameters shown in the first 
column of Table \ref{tab:tab1}, the value of $R_t$ is (the interquartile
range in parenthesis)
\[R_t = 1.4848~ (1.4798--1.4903). \] 
We calculated $R_t$ based on the 500 sets of parameter
values and then calculated the median and interquartile range.
When the metapopulation model is used without mobility ($\theta=0$), 
then the same expression,  Eq. (\ref{eq:R01node}), applies with 
the  parameters of the second column yielding 
\[R_t = 1.5972~ (1.5945--1.5983).\]
Finally, for the entire metapopulation network, $R_t$ is calculated as 
follows. We define the relevant vectors, focusing on the infectious/infected 
compartments ($E_i$, $A_i$, $I_i$) and ignoring the rest 
($S_i$, $U_i$, $R_i$, $D_i$):
\begin{align*}
\mathcal{F} =\left(
\begin{array}{c}
\frac{\beta_{SA}}{N_1} S_1 A_1 + \frac{\beta_{SI}}{N_1} S_1 I_1 \\
0 \\ 0\\ 
\vdots \\
\frac{\beta_{SA}}{N_8} S_8 A_8 + \frac{\beta_{SI}}{N_8} S_8 I_8 \\
0 \\ 0\\ 
\end{array}
\right),~~
\mathcal{V} = \left(
\begin{array}{c}
(\kappa_A +\kappa_I) E_1 - \theta \left( \sum_j M_{1j} \frac{E_j}{N_j} -   
\sum_j M_{j1} \frac{E_1}{N_1} \right)\\
-\kappa_A E_1 +\gamma_A A_1 - \theta \left( \sum_j M_{1j} \frac{A_j}{N_j} -   
\sum_j M_{j1} \frac{A_1}{N_1} \right)\\
-\kappa_I E_1 + (\kappa_R +\kappa_D) I_1 - \theta \left( \sum_j M_{1j} \frac{I_j}{N_j} -   
\sum_j M_{j1} \frac{I_1}{N_1} \right)\\ 
\vdots\\
(\kappa_A +\kappa_I) E_8 - \theta \left( \sum_j M_{8j} \frac{E_j}{N_j} -   
\sum_j M_{j8} \frac{E_8}{N_8} \right)\\
-\kappa_A E_8 +\gamma_A A_8 - \theta \left( \sum_j M_{8j} \frac{A_j}{N_j} -   
\sum_j M_{j8} \frac{A_8}{N_8} \right)\\
-\kappa_I E_8 + (\kappa_R +\kappa_D) I_8 - \theta \left( \sum_j M_{8j} \frac{I_j}{N_j} -   
\sum_j M_{j8} \frac{I_8}{N_8} \right)
\end{array}
\right)
\end{align*}

We then find  the Jacobian matrices of $\mathcal{F}, \mathcal{V}$ with 
respect to $E_i, A_i, I_i$ in  the order in which they appear. 
This yields 
two $24 \times 24$ matrices of the form:
\begin{align*}
&F= \left(
\begin{array}{ccccc}
F_{11} & O_{3 \times 3} & O_{3 \times 3} & \dots & O_{3 \times 3}\\
O_{3 \times 3} & F_{22} & O_{3 \times 3} & O_{3 \times 3} & 
O_{3 \times 3} \\
O_{3 \times 3}&  O_{3 \times 3}& F_{33} & O_{3 \times 3} & 
O_{3 \times 3} \\
\vdots & \vdots & \vdots &\ddots  & \vdots \\
O_{3 \times 3} &  O_{3 \times 3} & O_{3 \times 3}  & \dots & 
F_{88}
\end{array}
\right), ~~~ 
F_{ii} = \left( \begin{array}{ccc}
0 & \beta_{AS} \frac{S_i}{N_i} & \beta_{IS} \frac{S_i}{N_i} \\
0 & 0 &0 \\
0 & 0 &0 
\end{array} \right), ~~~ 
O_{3 \times 3} = \left( \begin{array}{ccc}
0 & 0 &0 \\
0 & 0 &0 \\
0 & 0 &0
\end{array} \right)
\\
&V=
\left(
\begin{array}{cccc}
V_{11} & V_{12} &  \dots & V_{18} \\
V_{21}&  V_{22} & \dots & V_{28} \\
\vdots &  \vdots &  \ddots & \vdots \\
V_{81} & V_{82} & \dots & V_{88}
\end{array}
\right), ~~~ V_{ii} =\left( \begin{array}{ccc}
\kappa_A +\kappa_I + \theta \sum_j \frac{M_{ji}}{Ni} & 0 & 0 \\
-\kappa_A  & \gamma_A + \theta \sum_j \frac{M_{ji}}{N_i} & 0 \\
-\kappa_I & 0 &  \kappa_R+\kappa_D + \theta \sum_j \frac{M_{ji}}{N_i}
\end{array} 
\right) \\
& V_{ij} =  \left( \begin{array}{ccc}
-\theta \frac{M_{ij}}{N_j} & 0 & 0 \\
0 & -\theta \frac{M_{ij}}{N_j} & 0 \\
0 & 0 & -\theta \frac{M_{ij}}{N_j}
\end{array} \right)
\end{align*}

The reproduction number is the spectral radius of $F V^{-1}$ which in our case
has the value
\[R_t = 1.3349~ (1.2806-1.4581).\]
This is exactly the same value one obtains when using Eq. (\ref{eq:R01node})
if one completely disregards the mobility terms in matrix $V$, i.e., 
with the parameter values of  the third  column in Table \ref{tab:tab1}. 
Hence, the change 
in $R_t$ is due to the different values in third column of Table 
\ref{tab:tab1}, and not due to the terms containing $M_{ij}$ 
in matrix $V$. In other words, the effect of mobility is to change
the estimated parameters, which while they 
do not alter the $R_t$ {(which, as mentioned earlier
is effectively  evaluated at the first day of our simulations)},
they have a large effect on  the dynamics later. When mobility is included in the model, 
the interquartile intervals of each parameter value are significantly wider,
which is also reflected in the corresponding  interval for $R_t$.

\section{Discussion-conclusion}

In the present work, we revisited the formulation of metapopulation
models, motivated by the interest towards describing a ``relatively
small'' region (the autonomous community of Andalusia within Spain)
with well-defined and available in a time-resolved
manner data regarding the mobility across provinces. It is also
a region without
an extensive influx (or outflux) of populations, e.g., through major
international airport hubs. This appears to render this case
a fertile ground for the application of metapopulation models. 

In that vein, in addition to a prototypical model for each node, 
involving susceptibles, exposed, asymptomatic and symptomatically infected,
as well as recovered from each of these categories and fatalities,
we considered different possibilities on how to incorporate human mobility
across the nodes. We explored the model for the entire autonomous
community of Andalusia (without sub-nodes), the model where the nodes
do not feature mobility between them (independent nodes) and the canonical
case proposed where mobility is incorporated.
One of the main findings of the present work is that in the absence of mobility among
nodes 
%($\theta=0$), 
the model is  unable to predict the wave of infections that took place 
in the fall of 2020. It has long been known that human mobility is crucial at the 
beginning stages of an epidemic, when the infection is seeded in various locations
\cite{chinazzi2020, kraemer2020, wesolowski2016}. It has also been noted that mobility 
may also affect contact rates which in turn affect disease transmission \cite{wesolowski2016}.
The present study suggests that population flows are critically important in periods 
during an epidemic when there are no restrictions on mobility.  Moreover, while 
there are numerous
ways of incorporating mobility, including gravity models, origin-destination matrices, etc., we believe that 
at present the optimal inclusion should be time-resolved.
Dynamical information stemming from mobile-phone data seamlessly 
incorporates aspects such as the weekly or seasonal variations of human mobility;
hence it captures more accurately the resulting increases or decreases
in the probability of formation of an epidemic wave of infection. 
However, when this is not possible, we also offer details on how
origin-destination matrices, obtained by the 
gravity-law can be calculated to be used in
a metapopulation model.

Nevertheless, we will certainly refrain from assigning full responsibility to human mobility for the wave of infections in the fall of 2020, or indeed more 
generally during the second wave of the pandemic.
It is clear that there exist numerous factors that may have contributed to the
relevant features, including, e.g., seasonality~\cite{danon2021}. It would be interesting to explore
these factors and their interplay with mobility further both in the
context of the second wave (as here) in other regions, but also as
concerns subsequent waves of the pandemic, where other key
factors, such as the existence and the role of vaccinations~\cite{vacc2021}
need to be taken into consideration. Such studies will be deferred to future
publications.

\acknowledgments
ZR, EK, SL, PGK, MB and GAK acknowledge support through the C3.ai Inc. and Microsoft Corporation. ZR also acknowledges support from the NSF through grant
 DMS-1815764. JCM acknowledges support from EU (FEDER program 2014-2020) through both Consejer\'{\i}a de Econom\'{\i}a, Conocimiento, Empresas y Universidad de la Junta de Andaluc\'{\i}a (under the projects P18-RT-3480 and US-1380977), and MCIN/AEI/10.13039/501100011033 (under the project  PID2020-112620GB-I00).

\clearpage

\section{Supporting Information} 
\label{sec:supporting_info}

\subsection{Overview}
The supplementary section describes the methodology leveraged for computing origin-destination (O-D) light-duty
vehicle highway travel patterns compatible with empirical data of coarse grained flows in
highway transportation networks. We propose the following approach to construct the O-D matrix: 
given population data from the Instituto Nacional de Estad{\'i}stica \cite{popdata}
and distances of city pairs we can apply the  gravity law. We present the method\sout{s} and comment on the use of
possible datasets that could serve as inputs to a metapopulation model.
Such insights enable comparison of the performance of the data-driven O-D matrix and
the rest of the methods typically applied for O-D estimation. Such a comparison
is presented in Fig.~\ref{fig:mobilcompar} in the main text.

%%%%%%%%%%%%%%%%%%%%%%%%%%%%%%%%%%%%%%%%%%%%%%%%%%%%%%%%%%%%%%%%%%%%%%%%%%%%%%%%

%\clearpage

\subsection{Methodology}
\subsubsection{Gravity law: Traffic counts}

Given the population of provinces and distances between the capital city of province pairs, the gravity model as described by
Eq.~(\ref{eq:gravity}) in the main text can be leveraged with the parameters presented in Table \ref{tab:gravity_parameters}. 
\begin{table}[!htb]
\centering
\caption{Parameters of gravity law \cite{balcan2009, stefanouli2017gravity}}
\begin{tabular}{ccc}
\toprule
$d$(km) & Parameter & Estimate\\
\hline
$\leq 300$ & $\alpha$ & 0.46\\ 
           & $\gamma$ & 0.64\\ 
           & $\beta$ & 0.0122\\ 
\textgreater 300 & $\alpha$ & 0.35\\
           & $\gamma$ & 0.37\\
           & $\beta$ & N/A\\
           & $C$     & 0.04289 \\
\bottomrule
\end{tabular}
\label{tab:gravity_parameters}
\end{table}
When the distance between an O-D pair ($ij$) is larger than 300 km, $\beta$ is set to be N/A, which denotes that the denominator will be approaching 0 and $w_{ij}$ is also approaching 0. 

The Origin-Destination matrix for the eight Andalusian provinces
as calculated from the gravity law with the parameters reported in
Table~\ref{tab:gravity_parameters} is presented in Table \ref{tab:gravity_od_sp}.
As mentioned in the main text this O-D matrix is static, and the data we used
referred to 2019-2020, i.e., before the pandemic started.

\subsection{Highway O-D matrix construction}

Table \ref{tab:gravity_od_sp} shows
{the daily number of trips} between O-D pairs ($ij$),
as calculated from the gravity law.
\begin{table}[htb]
\centering
\caption{O-D matrix based on the gravity law.}
\begin{tabular}{ccccccccc}
\toprule
   & Almeria & Cadiz & Cordoba & Granada & Huelva & Jaen & Malaga & Sevilea\\
\hline
Almeria & 0 & 0 & 0 & 4139 & 0 & 873 & 4863 & 0\\
Cadiz & 0 & 0 & 2391 & 0 & 2645 & 0 & 4704 & 21312\\
Cordoba & 0 & 2536 & 0 & 3323 & 1658 & 4211 & 10110 & 14025\\
Granada & 4018 & 0 & 3531 & 0 & 0 & 7051 & 13425 & 3253\\
Huelva & 0 & 3248 & 1920 & 0 & 0 & 0 & 0 & 17683\\
Jaen & 964 & 0 & 5092 & 8022 & 0 & 0 & 3728 & 2612\\
Malaga & 4015 & 4509 & 9138 & 11417 & 0 & 2787 & 0 & 9059\\
Seville & 0 & 19750 & 12257 & 2675 & 13346 & 1888 & 8758 & 0 \\
\bottomrule
\end{tabular}
\label{tab:gravity_od_sp}
\end{table}

\begin{comment}

Table~\ref{tab:max_entropy_od_sp} presents the constructed,
pre-pandemic, O-D matrix based on link flows obtained from the
{highway network} map and using maximum entropy theory.

\begin{table}[htb]
\centering
\caption{O-D matrix based on maximization entropy principles.}
\begin{tabular}{ccccccccc}
\toprule
   & Almeria & Cadiz & Cordoba & Granada & Huelva & Jaen & Malaga & Seville\\
\hline
Almeria & 0 & 0 & 5246 & 0 & 0 & 2645 & 2645 & 775\\
Cadiz & 0 & 0 & 0 & 0 & 7153 & 25858 & 0 & 11117\\
Cordoba & 5246 & 0 & 0 & 961 & 355 & 272 & 0 & 11117\\
Granada & 0 & 0 & 961 & 0 & 0 & 3 & 961 & 0\\
Huelva & 0 & 7153 & 355 & 0 & 0 & 0 & 27238 & 0\\
Jaen & 2645 & 25858 & 272 & 3 & 0 & 0 & 12035 & 314\\
Malaga & 2645 & 0 & 0 & 961 & 27238 & 12035 & 0 & 11117\\
Seville & 775 & 11117 & 11117 & 0 & 0 & 314 & 11117 & 0\\
\bottomrule
\end{tabular}
\label{tab:max_entropy_od_sp}
\end{table}

\end{comment}


\begin{thebibliography}{100}

%\bibitem{TrafficMapSpain2018}
%Traffic map.
%\newblock
%  https://cdn.mitma.gob.es/portal-web-drupal/Mapas\_de\_trafico/mapa\_imd\_2018.pdf.

\bibitem{ando21}
S. Ando, Y. Matsuzawa, H. Tsurui, T. Mizutani, D. Hall,
  and Y. Kuroda.
\newblock Stochastic modelling of the effects of human-mobility restriction and
  viral infection characteristics on the spread of covid-19.
\newblock {\em Scientific Reports}, 11(1):6856, 2021.

\bibitem{arenas2020}
A.~Arenas, W.~Cota, J.~Gomez-Garde\~nes, S.~Gomez, S.~Granell, J.~T. Matamalas,
  D.~Soriano-Panos, and B.~Steinegger.
\newblock Modeling the spatiotemporal epidemic spreading of {C}{O}{V}{I}{D}-19
  and the impact of mobility and social distancing interventions.
\newblock {\em Physical Review X}, 10:041055, 2020.

\bibitem{badr2020association}
H.S. Badr, H. Du, M. Marshall, E. Dong, M.M. Squire,
  and L.M. Gardner.
\newblock Association between mobility patterns and {C}{O}{V}{I}{D}-19
  transmission in the {U}{S}{A}: a mathematical modelling study.
\newblock {\em The Lancet Infectious Diseases}, 20(11):1247--1254, 2020.

\bibitem{balcan2009}
D.~Balcan, V.~Colizza, B.~Goncalves, H.~Hu, J.~J. Ramasco, and A.~Vespignani.
\newblock Multiscale mobility networks and the spatial spreading of infectious
  diseases.
\newblock {\em Proceedings of the National Academy of Sciences of the USA},
  106:21484--21489, 2009.

\bibitem{brockmann2011}
V.~Belik, T.~Geisel, and D.~Brockmann.
\newblock Natural human mobility patterns and spatial spread of infectious
  diseases.
\newblock {\em Physical Review X}, 1:011001, 2011.

\bibitem{Calleri2021}
F. Calleri, G. Nastasi, and V. Romano.
\newblock Continuous-time stochastic processes for the spread of covid-19
  disease simulated via a monte carlo approach and comparison with
  deterministic models.
\newblock {\em Journal of Mathematical Biology}, 83(4):1--26, 2021.

\bibitem{review_meta}
D. Calvetti, A.P. Hoover, J. Rose, and E. Somersalo.
\newblock Metapopulation network models for understanding, predicting, and
  managing the coronavirus disease {C}{O}{V}{I}{D}-19.
\newblock {\em Frontiers in Physics}, 8:261, 2020.

\bibitem{chen2014modeling}
D. Chen.
\newblock Modeling the spread of infectious diseases: A review.
\newblock {\em in: Analyzing and modeling spatial and temporal dynamics of
  infectious diseases}, pages 19--42, 2014.

\bibitem{chinazzi2020}
M. Chinazzi, J.T. Davis, M. Ajelli, C. Gioannini, M.
  Litvinova, S. Merler, A. Pastore~y~Piontti, K. Mu, L. Rossi,
  K. Sun, C. Viboud, X. Xiong, H. Yu, M.E. Halloran,
  I.M. Longini, and A. Vespignani.
\newblock The effect of travel restrictions on the spread of the 2019 novel
  coronavirus ({C}{O}{V}{I}{D}-19) outbreak.
\newblock {\em Science}, 368:395--400, 2020.

\bibitem{costa2020}
G.~S. Costa, W.~Cota, and S.~C. Ferreira.
\newblock Outbreak diversity in epidemic waves propagating through distinct
  geographical scales.
\newblock {\em Physical Review Research}, 2:043306, 2020.

\bibitem{cuevas2021}
J.~Cuevas-Maraver, P.G. Kevrekidis, Q.Y. Chen, G.A. Kevrekidis, Z.~Rapti, and
  Y.~Drossinos.
\newblock Lockdown measures and their impact on single- and two-age-structured
  epidemic model for the {C}{O}{V}{I}{D}-19 outbreak in mexico.
\newblock {\em Mathematical Biosciences}, 336:108590, 2021.

\bibitem{danon2021}
L.~Danon, E.~Brooks-Pollock, M.~Bailey, and M.~Keeling.
\newblock A spatial model of {C}{O}{V}{I}{D}-19 transmission in {E}ngland and
  {W}ales: early spread, peak timing and the impact of seasonality.
\newblock {\em Philosophical Transaction of the Royal Society B}, 376:20200272,
  2021.

\bibitem{danon2009}
L.~Danon, T.~House, and M.J. Keeling.
\newblock The role of routine versus random movements on the spread of disease
  in {G}reat {B}ritain.
\newblock {\em Epidemics}, 1:250--258, 2009.

\bibitem{CNE}
Ministerio de~Ciencia~e Innovacion.
\newblock {C}{O}{V}{I}{D}-19 en {E}spana.
\newblock https://cnecovid.isciii.es/covid19/.

\bibitem{de2011modelling}
J.~Ort{\'u}zar and L.G. Willumsen.
\newblock {\em Modelling transport}.
\newblock John Wiley \& Sons, 2011.

\bibitem{popdata}
Instituto~Nacional de~Estad{\'i}stica.
\newblock https://www.ine.es/en/.

\bibitem{diekmann1990}
O.~Diekmann, J.~A.~P. Heesterbeek, and J.~A.~J. Metz.
\newblock On the definition and the computation of the basic reproduction ratio
  ${R}_0$ in models for infectious diseases in heterogeneous populations.
\newblock {\em Journal of Mathematical Biology}, 28:365--382, 1990.

\bibitem{erlander1990gravity}
S. Erlander and N.F. Stewart.
\newblock {\em The gravity model in transportation analysis: theory and
  extensions}, volume~3.
\newblock Vsp, 1990.

\bibitem{glaeser2022}
E.~L. Glaeser, G.~Gorback, and S.~J. Redding.
\newblock Jue {I}nsight: {H}ow much does {C}{O}{V}{I}{D}-19 increase with
  mobility? {E}vidence from {N}ew {Y}ork and four other {U}.{S}. cities.
\newblock {\em Journal of Urban Economics}, 127:103292, 2022.

\bibitem{gomez2019}
S.~Gomez, A.~Fernandez, S.~Meloni, and A.~Arenas.
\newblock Impact of origin-destination information in epidemic spreading.
\newblock {\em Scientific Reports}, 9:2315, 2019.

\bibitem{holmdahl2020}
I.~Holmdahl and C.~Buckee.
\newblock Wrong but useful- what {C}{O}{V}{I}{D}-19 epidemiological models can
  and cannot tell us.
\newblock {\em The New England Journal of Medicine}, 383(4):303--305, 2020.

\bibitem{keeling2000}
M.~J. Keeling and C.~A. Gilligan.
\newblock Metapopulation dynamics of bubonic plague.
\newblock {\em Nature}, 407:903--906, 2000.

\bibitem{keeling2002}
M.~J. Kelling and P.~Rohani.
\newblock Estimating spatial coupling in epidemiological systems: a mechanistic
  approach.
\newblock {\em Ecology Letters}, 5:20--29, 2002.

\bibitem{george2022}
G.~A. Kevrekidis, Z.~Rapti, Y.~Drossinos, P.G. Kevrekidis, M.~A. Barmann, Q.~Y.
  Chen, and J.~Cuevas-Maraver.
\newblock Backcasting {C}{O}{V}{I}{D}-19: A physics-informed estimate for early
  case incidence.
\newblock https://arxiv.org/pdf/2202.00507.pdf, 2022.

\bibitem{kevrekidis2021}
P.~G. Kevrekidis, J.~Cuevas-Maraver, Y.~Drossinos, Z.~Rapti, and G.~A.
  Kevrekidis.
\newblock Reaction-diffusion spatial modeling of {C}{O}{V}{I}{D}-19: Greece and
  {A}ndalusia as case examples.
\newblock {\em Physical Review E}, 104:024412, 2021.

\bibitem{kraemer2020}
M.U.G. Kraemer, C.-H. Yang, B. Gutierrez, C.-H. Wu, B.
  Klein, D.M. Pigott, L. du~Plessis, N.R. Faria, R. Li, W.P. Hanage,
  J.S. Brownstein, M. Layan, A. Vespignani, H. Tian, C. Dye, O.G. Pybus, and S.V. Scarpino.
\newblock The effect of human mobility and control measures on the
  {C}{O}{V}{I}{D}-19 epidemic in {C}hina.
\newblock {\em Science}, 368:493--497, 2020.

\bibitem{li2020}
R.~Li, S.~Pei, B.~Chen, Y.~Song, T.~Zhang, W.~Yang, and J.~Shaman.
\newblock Substantial undocumented infections facilitates the rapid
  dissemination of novel coronavirus ({S}{A}{R}{S}-{C}o{V}-2).
\newblock {\em Science}, 368:489--493, 2020.

\bibitem{mammeri2020}
Y.~Mammeri.
\newblock A reaction-diffusion system to better comprehend the unlockdown:
  Application of {S}{E}{I}{R}-type model with diffusion to the spatial spread
  of {C}{O}{V}{I}{D}-19 in {F}rance.
\newblock {\em Computational and Mathematical Biophysics}, 8:102--113, 2020.

\bibitem{mccallum2001should}
H. McCallum, N. Barlow, and J. Hone.
\newblock How should pathogen transmission be modelled?
\newblock {\em Trends in ecology \& evolution}, 16(6):295--300, 2001.

\bibitem{MITMA}
Movilidad y Agenda~Urbana Ministerio~de Transportes.
\newblock Estudio de movilidad con big data.
\newblock
  https://www.mitma.gob.es/ministerio/covid-19/evolucion-movilidad-big-data.

\bibitem{pei2018}
S.~Pei, S.~Kandula, W.~Yang, , and J.~Shaman.
\newblock Forecasting the spatial transmission of influenza in the {U}nited
  {S}tates.
\newblock {\em Proceedings of the National Academy of Sciences of the USA},
  115:2752--2757, 2018.

\bibitem{bhatta2020}
M.~Peirlinck, K.~Linka, F.~Sahli~Costabal, J.~Bhattacharya, E.~Bendavid,
  J.~A.~P. Ioannidis, and E.~Kuhl.
\newblock Visualizing the invisible: The effect of asymptomatic transmission on
  the outbreak dynamics of {C}{O}{V}{I}{D}-19.
\newblock {\em Computer Methods in Applied Mechanics and Engineering}, 2020.

\bibitem{prasse2022}
B.~Prasse, M.~A. Achterberg, and P.~V. Mieghem.
\newblock Accuracy of predicting epidemic outbreaks.
\newblock {\em Physical Review E}, 105:014302, 2022.

\bibitem{schlapfer2021}
M.~Schlapfer, L.~Dong, K.~O'Keeffe, P.~Santi, M.~Szell, H.~Salat,
  S.~Anklesaria, M.~Vazifeh, C.~Ratti, and G.~B. West.
\newblock The universal visitation law of human mobility.
\newblock {\em Nature}, 593:522--540, 2021.

\bibitem{simini2012}
F.~Simini, M.~C. Gonzalez, , A.~Maritan, and A.-L. Barabasi.
\newblock A universal model for mobility and migration patterns.
\newblock {\em Nature}, 484:96--100, 2012.

\bibitem{stefanouli2017gravity}
M. Stefanouli and S. Polyzos.
\newblock Gravity vs radiation model: two approaches on commuting in {G}reece.
\newblock {\em Transportation research procedia}, 24:65--72, 2017.

\bibitem{tizzoni2014}
M.~Tizzoni, P.~Bajardi, Decuyper A., G.~K. Kam~King, C.~M. Schneider,
  V.~Blondel, Z.~Smoreda, M.~C. Gonzalez, and V.~Colizza.
\newblock On the human mobility proxies for modeling epidemics.
\newblock {\em PLoS Computational Biology}, 10:e1003716, 2014.

\bibitem{vacc2021}
T.~Usherwood, Z.~LaJoie, and V.~Srivastava.
\newblock A model and predictions for {C}{O}{V}{I}{D}-19 considering population
  behavior and vaccination.
\newblock {\em Scientific Reports}, 11:12051, 2021.

\bibitem{van1980most}
H.J. Van~Zuylen and L.G. Willumsen.
\newblock The most likely trip matrix estimated from traffic counts.
\newblock {\em Transportation Research Part B: Methodological}, 14(3):281--293,
  1980.

\bibitem{viboud2006}
C.~Viboud, O.~N. Bjornstad, D.~L. Smith, L.~Simonsen, M.~A. Miller, and B.~T.
  Grenfell.
\newblock Synchrony, waves, and spatial hierarchies in the spread of influenza.
\newblock {\em Science}, 312:447--451, 2006.

\bibitem{viguerie2021}
A.~Viguerie, G.~Lorenzo, F.~Auricchio, D.~Baroli, T.~J.~R. Hughes, A.~Patton,
  A.~Reali, T.~E. Yankeelov, and A.~Veneziani.
\newblock Simulating the spread of {C}{O}{V}{I}{D}-19 via a spatially-resolved
  susceptible–exposed–infected–recovered–deceased ({S}{E}{I}{R}{D})
  model with heterogeneous diffusion.
\newblock {\em Applied Mathematics Letters}, 111:106617, 2021.

\bibitem{galvani2020}
C.~R. Wells, P.~Sah, S.~M. Moghadas, A.~Pandey, A.~Shoukat, Y.~Wang, Z.~Wang,
  L.~A. Meyers, B.~H. Singer, and A.~P. Galvani.
\newblock Impact of international travel and border control measures on the
  global spread of the novel 2019 coronavirus outbreak.
\newblock {\em Proceedings of the National Academy of Sciences of the USA},
  117:7504--7509, 2020.

\bibitem{wesolowski2016}
A.~Wesolowski, C.~O. Buckee, K.~Engo-Mongen, and C.~J.~E. Metcalf.
\newblock Connecting mobility to infectious diseases: the promise and limits of
  mobile phone data.
\newblock {\em The Journal of Infectious Diseases}, 214:S414--420, 2016.

\bibitem{willumsen1981}
L.~G. Willumsen.
\newblock Simplified transport models based on traffic counts.
\newblock {\em Transportation}, 10:257--278, 1981.

\bibitem{xia2004}
Y.~Xia, O.~N. Bjornstad, and B.~T. Grenfell.
\newblock Measles metapopulation dynamics: A gravity model for epidemiological
  coupling and dynamics.
\newblock {\em The American Naturalist}, 164(2):267--281, 2004.

\bibitem{yabe2020non}
T. Yabe, K. Tsubouchi, N. Fujiwara, T. Wada, Y. Sekimoto, and S.V. Ukkusuri.
\newblock Non-compulsory measures sufficiently reduced human mobility in tokyo
  during the covid-19 epidemic.
\newblock {\em Scientific reports}, 10(1):1--9, 2020.

\bibitem{zipf1946}
G.~K. Zipf.
\newblock The {P}1 {P}2/{D} hypothesis: on the intercity movement of persons.
\newblock {\em American Sociological Review}, 11:677--686, 1946.

\end{thebibliography}
\end{document}